\documentclass[12pt,preprint]{aastex}
\begin{document}

\newcommand{\dif}{\mathrm{d}}
\newcommand{\mclust}{\texttt{mclust}}
\newcommand{\optim}{\texttt{optim}}
\newcommand{\sparr}{\texttt{sparr}}
\newcommand{\spatstat}   {\texttt{spatstat}}
\newcommand{\mlpowerlaw}   {\texttt{ml\_powerlaw}}

\newcommand{\todo}[1]       {\noindent[{\bf \footnotesize #1}]} \newcommand{\tbr}[1]        {#1}
\newcommand{\tbd}           {\_\_\_\_  {\bf (TBD)}}     \newcommand{\revise}[1]     {{#1}}

\newcommand{\mystix} {MYStIX}
\newcommand{\Spitzer} {{\em Spitzer}}
\newcommand{\Herschel} {{\em Herschel}}
\newcommand{\Chandra} {{\em Chandra}}
\newcommand{\ACIS}    {{ACIS}}
\newcommand{\CIAO}    {{\em CIAO}}
\newcommand{\Sherpa}  {{\em Sherpa}}
\newcommand{\Chart}   {{\em ChaRT}}
\newcommand{\SAOTrace}{{\em SAOTrace}}
\newcommand{\DSnine}  {{\em DS9}}
\newcommand{\MARX}    {{\em MARX}}
\newcommand{\AEacro}  {{\em AE}}
\newcommand{\TARA}    {{\em TARA}}
\newcommand{\XSPEC}   {{\em XSPEC}}
\newcommand{\FTOOLS}  {{\em FTOOLS}}
\newcommand{\HEASOFT} {{\em HEASOFT}}
\newcommand{\IDL}     {{\em IDL}}
\newcommand{\Rlan}   {{\em R}}

\shorttitle{Evolutionary States of Young Stellar Clusters}
\shortauthors{Kuhn et al.}
\slugcomment{Accepted for publication in ApJ, July 2015}

\title{The Spatial Structure of Young Stellar Clusters. III. Physical Properties and Evolutionary States}

\author{Michael A. Kuhn\altaffilmark{1,2,3}, Eric D. Feigelson\altaffilmark{1}, Konstantin V. Getman\altaffilmark{1}, Alison Sills\altaffilmark{4}, Matthew R. Bate\altaffilmark{5}, Jordanka Borissova\altaffilmark{2,3}} 
\altaffiltext{1}{Department of Astronomy \& Astrophysics, 525 Davey Laboratory, Pennsylvania State University, University Park, PA 16802, USA}
\altaffiltext{2}{Instituto de Fisica y Astronom\'{i}a, Universidad de Valpara\'{i}so, Gran Breta\~{n}a 1111, Playa Ancha, Valpara\'{i}so, Chile}
\altaffiltext{3}{Millennium Institute of Astrophysics, MAS}
\altaffiltext{4}{Department of Physics, McMaster University, 1280 Main Street West, Hamilton, ON L8S 4M1, Canada}
\altaffiltext{5}{Department of Physics and Astronomy, University of Exeter, Stocker Road, Exeter, Devon EX4 4SB, UK}

\begin{abstract}
We analyze the physical properties of stellar clusters that are detected in massive star-forming regions in the \mystix\ project---a comparative, multiwavelength study of young stellar clusters within 3.6~kpc that contain at least one O-type star. Tabulated properties of subclusters in these regions include physical sizes and shapes, intrinsic numbers of stars, absorptions by the molecular clouds, and median subcluster ages. Physical signs of dynamical evolution are present in the relations of these properties, including statistically significant correlations between subcluster size, central density, and age, which are likely the result of cluster expansion after gas removal. We argue that many of the subclusters identified in Paper~I are gravitationally bound because their radii are significantly less than what would be expected from freely expanding clumps of stars with a typical initial stellar velocity dispersion of $\sim$3~km~s$^{-1}$ for star-forming regions. We explore a model for cluster formation in which structurally simpler clusters are built up hierarchically through the mergers of subclusters---subcluster mergers are indicated by an inverse relation between the numbers of stars in a subcluster and their central densities (also seen as a density vs.\ radius relation that is less steep than would be expected from pure expansion). We discuss implications of these effects for the dynamical relaxation of young stellar clusters.\end{abstract}

\section{Introduction}

Stars form from hierarchically collapsing molecular clouds, which leads to clustered star-formation often reflecting the structure of the molecular cloud. Observations of massive star-forming regions (MSFRs) at different stages of their star-forming lifetime (typically $<$5--10~Myr), reveal diverse stellar cluster structure, which provides information about cluster formation, cluster dynamics, and future cluster survival or destruction---problems that have wider implications for the origin of stellar populations in galaxies \citep[e.g.,][]{Lada03,PortegiesZwart10}. Important questions on these topics include:
\begin{description}
\item[How are stellar clusters formed?] Do they freeze out of molecular clouds in a single crossing time \citep{Elmegreen00,VazquezSemadeni03}? Are they built up by gradual star formation over many crossing times \citep{Tan06}? Do they form from the inside out \citep{Pfalzner11}? Or, are clusters formed from the hierarchical merger of subclusters \citep{McMillan07,Maschberger10,Bonnell03,Bate09a}?
\item[What is the origin of stellar clusters with different properties?] Do stellar clusters start from extremely dense states with $\sim$10$^6$~stars~pc$^{-3}$ \citep{Bate09a,Moeckel10}? Are observed differences in Galactic open cluster properties due to cluster evolution \citep{Pfalzner09,Gieles12}? 
Is the mass segregation that is seen in some regions primordial, indicating different star-formation mechanisms for high and low mass stars, or can it be explained dynamically \citep{Bonnell98,Allison09}?
\item[What environments lead to bound stellar clusters after the end of star for\-ma\-tion?] What roles do star-formation efficiency, cluster dynamics, star-formation feedback effects, and the tidal disruption of molecular clouds play \citep{Kruijssen12}? 
\end{description}

Advancements in multiwavelength studies of young stellar populations in Galactic star-forming regions, for example the \mystix\ \citep{Feigelson13} and MOXC \citep{Townsley14} projects, have led to better censuses of young stars \citep[both high and low-mass; both disk-bearing and disk-free;][]{Broos13} that reveal snapshots of the structure of young stellar clusters \citep[][henceforth Paper~I]{Kuhn14a}. The complex cluster structures seen in these regions resemble structures produced by hydrodynamical simulations of star-cluster formation from turbulent molecular clouds \citep[e.g.,][]{Bate03,Bate09a,Bonnell11,Dale12,Walch12}.
Although it is not possible to watch star-cluster formation and evolution unfold in real clusters in the Galaxy, through comparison of the properties of a large sample of clusters spanning a range of environments and ages, it is possible to make inferences about processes such as gas removal, cluster expansion, subcluster mergers, dynamical boundedness and relaxation, and cluster dispersal, which all affect the spatial structure of clusters.
\citet{Getman14} provide age estimates for over 100 subclusters in \mystix\ MSFRs, which span a wide age range between 0.5 and 5~Myr, showing that individual regions often have spatially segregated structures with different ages. The age information is used here to examine how subclusters dynamically evolve.

The organization of this paper is as follows. Section~2 describes the data that are available for stars and stellar subclusters from the \mystix\ project. Section~3 uses multivariate statistics to compare measured physical properties of subclusters. Section~4 derives dynamical properties for the subclusters and uses multivariate statistics to investigate their dynamical evolution. In the conclusion to this paper (Section~5) we discuss implications for cluster-formation theory. 

\section{Data}

This paper uses the Paper~I catalog of 142 subclusters of young stars in 17 \mystix\ star-forming regions. These subclusters were found and characterized using the stars in the \mystix\ Probable Complex Member (MPCM) catalogs from \citet{Broos13}. The multiwavelength data analysis efforts that went into this catalog are described by \citet{Feigelson13}, \citet{Kuhn13a}, \citet{Townsley14}, \citet{King13}, \citet{Kuhn13b}, \citet{Naylor13}, and \citet{Povich13}, which provided uniform data coverage across the 17 star-forming regions investigated here, including the most comprehensive and reliable lists of young stars in many of the nearest MSFRs. For the analysis of spatial structure, we use a subset of the MPCM sources from which weak X-ray sources have been removed in order to produce a sample with spatially uniform X-ray flux detection limits (Paper~I).

For cluster analysis we used the ``finite mixture model'' method. To construct these models, the projected spatial distribution of stars in each individual subcluster is modeled with its own probability density, which has the form of an ``isothermal ellipsoid'' (Equation~4 in Paper~I). The isothermal ellipsoid is roughly constant within a core radius, $r_c$, but decreases with a -2 power law-index outside this radius. Each subcluster is described by six parameters: cluster center (R.A.,~Dec.), subcluster core radius\footnote{The radius $r_c$ is the harmonic mean of the semi-major and semi-minor axes of the core ellipse.} ($r_c$), ellipticity ($\epsilon$), orientation ($\phi$), and central surface density ($\Sigma_0$). In this case, the finite mixture model is surface-density distribution obtained through summation of every isothermal ellipsoid for all subclusters plus a model component for unclustered stars (Equation~5 in Paper~I), and the likelihood of a particular finite mixture model given a set of points can be calculated using Equation~6 in Paper~I. The best-fitting parameters for each subcluster are found using maximum likelihood estimation, and the number of subclusters is determined using the Akeike Information Criterion, a likelihood that is penalized by the total number of parameters in the model. We find good agreement between stellar surface density estimate from this parametric model and surface densities estimated using an adaptive smoothing algorithm.

\citet[][henceforth Paper~II]{Kuhn14b} provides estimates of the total number of stars in each subcluster. The number of low-mass stars missing from the MPCM catalogs is inferred from the X-ray luminosity function \citep[e.g.,][]{Getman05,Feigelson05} and initial mass function, which provide independent but consistent estimates of intrinsic populations.\footnote{We choose to use the estimates obtained via the X-ray luminosity function because there are fewer sources of systematic uncertainty in estimated X-ray luminosities compared to estimated stellar masses used by Paper~II.} Since Paper~I does not provide outer truncation radii for subclusters, we must chose a characteristic radius to describe the size of clusters. Here, we define $r_4 = 4 r_c$ as a characteristic radius---this coincides with our radius choice for determining subcluster assignments for individual stars in Paper~I and would be roughly the projected half-mass radius for a subcluster with an outer truncation radius at $r_t\sim17r_c$. We define $n_4$ to be the number of stars within a projected radius $r_4$; and, following Equation~3 of Paper~I, $n_4$, $r_4$, and $\Sigma_0$ are related to each other by the equation
\begin{equation}\label{n4.eqn}
n_4 = 0.56\, r_4^2 \, \Sigma_0.
\end{equation}
The conversion between the central volume density and central surface density of an isothermal sphere is 
\begin{equation}\label{rho0.eqn}
\rho_0 = \Sigma_0/2\, r_c = 2\,\Sigma_0/r_4.
\end{equation}
And, we use the above equation as the definition of the parameter $\rho_0$ tabulated in this paper---this relation between central volume density and projected surface density would be approximately true for the isothermal ellipsoid if the subcluster is neither much more compressed nor extended along the line of sight than in the plane of the sky.\footnote{We cannot avoid this assumption because we do not have measurements of the small differences in distance to different cluster members. Similar approximations have been made by other studies of cluster central densities \citep[e.g.,][]{Pfalzner09}. } 
As a consequence of these definitions, any two of the variables $r_4$, $n_4$, $\Sigma_0$, and $\rho_0$ that we use to describe the modeled ellipsoids are sufficient to fully describe the model.

The obscuration of subclusters is measured using the near-IR $J-H$ color index and the X-ray median energy ({\it ME}) indicator. Paper~I provides median $J-H$ and {\it ME} values for the MPCMs assigned to each subcluster (larger values indicate higher obscuration). Although, obscuration is a projection effect, subclusters with greater obscuration tend to be more deeply embedded in the molecular cloud. 

Subcluster ages are obtained by \citet{Getman14} using the novel $Age_{JX}$ method, in which X-ray luminosities of low-mass pre-main sequence stars are used as a proxies for stellar masses and dereddened $J$-band luminosities as proxies for bolometric luminosities. Ages for individual young stars may have large uncertainty, due to statistical error in luminosities, uncertainties in dereddening, and the inherent scatter in the X-ray luminosity$\sim$mass relation. Nevertheless, we are not so much interested in exact ages for individual objects as we are in consistent treatment that permits comparison between subclusters. Ages are calculated using the \citet{Siess97} PMS stellar evolution tracks---different evolutionary models may systematically shift ages, but are unlikely to change the ordering of the median stellar ages of different subclusters. 

Paper~I defined four heuristic classes of morphological structures (i.e., the arrangements of subclusters) seen in MSFRs, which include ``simple'' structures composed of an isolated isothermal ellipsoid, ``core-halo'' structure, ``clumpy'' structures, and long ``chains'' of subclusters. We label the subclusters by the class of the large-scale structure that they are a component of. For example, the star-forming region DR~21 has a chain structure, in which a line of subclusters is embedded in a dense molecular cloud filament, so all DR~21 subclusters (A through I) are labeled with ``chain.'' On the other hand, some regions show multiple structures; for example, the Carina region has an overall chain-like structure, but the Tr~14 cluster (our subclusters B and C) have a core-halo formation (Paper~I, their Figure 2). These classifications are based on the subclusters' placements relative to other subclusters in a region, rather than on the subclusters' own intrinsic properties.

Table~\ref{intrinsic.tab} presents the astrophysical properties of subclusters: the morphological class that a subcluster belongs to, subcluster size out to four core radii ($r_4$), the number of stars projected within four core radii ($n_4$), the central density of stars within a subcluster ($\sigma_0$ and $\rho_0$), subcluster ellipticity ($\epsilon$), median interstellar medium absorption indicators ($J-H$ and $ME$), and median stellar age ($Age_{JX}$). Logarithmic values are used for radii, numbers of stars, densities, and ages due to the large dynamic range of their values. We preserve two significant figures beyond the decimal point for the logarithmic values to preserve the peaks in the distributions when un-logged values are used; however, uncertainties on radius, number of stars, and age are nearer 0.1~dex ($\approx$25\%).

In many of the scatter plots in this paper, characteristic error bars are shown as $+$ symbols representing the median statistical uncertainties calculated for subcluster properties. 
The Fisher information matrices for the maximum-likelihood cluster models provide estimates for uncertainty on $r_4$ and $\epsilon$. These estimates do not account for systematic uncertainties in model selection but are useful for estimating the general effect of uncertainty on measurements. The uncertainties reported for $n_4$ come from the scatter in number of observed stars when drawing from a Poisson distribution, scaled to the inferred number of stars. The $n_4$ uncertainties (listed in Paper~II, their Table~2) do not take into account the systematic errors in estimating total stellar populations from observed numbers of stars that arise from both the extrapolation using the X-ray luminosity function and the systematic uncertainty in subcluster model. Uncertainties in $r_4$ and $n_4$ are propagated to $\Sigma_0$ and $\rho_o$, assuming that uncertainty is normally distributed. Uncertainty on the medians of $J-H$, $ME$, and $Age_{JX}$ are calculated via bootstrap resampling \citep[See][their Table~2]{Getman14}. The resulting typical uncertainties are 0.091~dex for $r_4$, 0.25 for $\epsilon$, 0.070~dex for $n_4$,  0.19~dex for $\Sigma_0$, 0.283~dex for $\rho_0$, 0.093~dex for $Age_{JX}$, 0.057~mag for median~$J-H$, and 0.073~keV for median~$ME$. 
Nevertheless, we estimate scatter in regressions between subcluster properties empirically from the data.

\clearpage\clearpage

\begin{deluxetable}{llrrrrrrrr}
\tablecaption{Intrinsic Subcluster Properties \label{intrinsic.tab}}
\tabletypesize{\tiny}\tablewidth{0pt}
\tablehead{
\colhead{Subcluster} & \colhead{Morph.} &\colhead{$\log r_4$} & \colhead{$\log N_4$} & \colhead{$\log \Sigma_0$} & \colhead{$\log \rho_0$} & \colhead{$\epsilon$} & \colhead{$J-H$} & \colhead{$ME$} & \colhead{$\log$ Age} \\
\colhead{} & \colhead{Class} &\colhead{(pc)} & \colhead{(stars)} & \colhead{(stars pc$^{-2}$)} & \colhead{(stars pc$^{-3}$)} & \colhead{} & \colhead{(mag)} & \colhead{(keV)} & \colhead{(yr)}\\
\colhead{(1)} & \colhead{(2)} & \colhead{(3)} & \colhead{(4)} & \colhead{(5)} & \colhead{(6)}& \colhead{(7)} & \colhead{(8)}& \colhead{(9)}& \colhead{(10)}
}
\startdata
orion~B & core & -0.71 & 2.17 & 3.85 & 4.86 & 0.3 & 0.87 & 1.6 & 6.04 \\
orion~C & halo & -0.06 & 3.21 & 3.58 & 3.94 & 0.49 & 1.05 & 1.6 & 6.18 \\
orion~D & clumpy & -0.44 & 1.94 & 3.07 & 3.81 & 0.84 & 1.17 & 1.4 & 6.43 \\
flame~A & simple & -0.31 & 2.74 & 3.61 & 4.22 & 0.37 & 1.79 & 2.8 & 5.90 \\
w40~A & simple & -0.19 & 2.48 & 3.11 & 3.60 & 0.04 & 2.10 & 2.5 & 5.90 \\
rcw36~A & halo & -0.24 & 2.73 & 3.46 & 4.00 & 0.33 & 1.63 & 2.3 & 5.95 \\
rcw36~B & core & -1.10 & 1.66 & 4.12 & 5.51 & 0.85 & 2.14 & 2.8 & \nodata \\
ngc2264~A & chain & -0.48 & 1.21: & 2.42: & 3.20: & 0.14 & 0.83 & 1.1 & 5.95 \\
ngc2264~B & chain & -0.95 & 1.12: & 3.28: & 4.53: & 0.44 & 0.60 & 1.1 & \nodata \\
ngc2264~C & chain & -0.94 & 1.26: & 3.39: & 4.63: & 0.11 & 0.73 & 1.1 & \nodata \\
ngc2264~D & chain & -0.32 & 1.38 & 2.27 & 2.89 & 0.11 & 0.65 & 1.1 & 6.50 \\
ngc2264~E & chain & -0.05 & 2.16 & 2.52 & 2.88 & 0.47 & 0.63 & 1.1 & 6.50 \\
ngc2264~F & chain & -0.70 & 1.51 & 3.15 & 4.15 & 0.54 & 0.61 & 1.0 & \nodata \\
ngc2264~G & chain & -0.48 & 1.76 & 2.98 & 3.77 & 0.31 & 2.11 & 2.3 & 6.18 \\
ngc2264~H & chain & -0.16 & 1.51 & 2.08 & 2.54 & 0.19 & 0.65 & 1.0 & \nodata \\
ngc2264~I & chain & -0.53 & 1.85 & 3.16 & 3.99 & 0.55 & 1.16 & 1.6 & 6.18 \\
ngc2264~J & chain & -0.25 & $>$1.94 & $>$2.64 & $>$3.19 & 0.25 & 1.22 & 1.7 & 6.20 \\
ngc2264~K & chain & -0.17 & 2.25 & 2.85 & 3.32 & 0.55 & 0.75 & 1.3 & 6.34 \\
ngc2264~L & chain & -0.93 & $>$1.33 & $>$3.44 & $>$4.66 & 0.13 & 1.84 & 3.5 & \nodata \\
ngc2264~M & chain & -0.52 & 1.69 & 2.98 & 3.79 & 0.32 & 0.73 & 1.2 & 6.08 \\
rosette~A & simple & 0.61 & 2.28 & 1.31 & 1.00 & 0.28 & 0.79 & 1.4 & \nodata \\
rosette~B & clumpy & -0.20 & 1.00: & 1.67: & 2.17: & 0.00 & 0.73 & 1.4 & 6.63 \\
rosette~C & clumpy & -0.20 & $>$1.26 & $>$1.85 & $>$2.36 & 0.71 & 0.77 & 1.4 & 6.61 \\
rosette~D & clumpy & -0.52 & 1.44 & 2.73 & 3.55 & 0.72 & 1.17 & 1.4 & \nodata \\
rosette~E & clumpy & 0.54 & 2.76 & 1.93 & 1.69 & 0.08 & 0.76 & 1.3 & 6.48 \\
rosette~F & clumpy & -0.27 & 0.90: & 1.68: & 2.25: & 0.51 & 0.83 & 1.4 & 6.60 \\
rosette~G & clumpy & -0.20 & \nodata & \nodata & \nodata & 0.00 & 0.76 & 1.0 & \nodata \\
rosette~H & clumpy & 0.13 & 1.70 & 1.69 & 1.86 & 0.82 & 0.84 & 1.3 & \nodata \\
rosette~I & clumpy & -0.17 & $>$1.11 & $>$1.62 & $>$2.09 & 0.56 & 0.94 & 1.6 & \nodata \\
rosette~J & clumpy & -0.32 & \nodata & \nodata & \nodata & 0.53 & 2.18 & 2.4 & \nodata \\
rosette~K & clumpy & -0.45 & \nodata & \nodata & \nodata & 0.00 & 2.00 & 2.6 & \nodata \\
rosette~L & clumpy & 0.51 & 2.46 & 1.69 & 1.48 & 0.51 & 1.10 & 1.5 & 6.43 \\
rosette~M & clumpy & 0.21 & 2.16 & 2.00 & 2.10 & 0.61 & 2.21 & 2.3 & 6.28 \\
rosette~N & clumpy & -0.15 & $>$1.50 & $>$1.98 & $>$2.44 & 0.08 & 1.62 & 1.9 & 6.11 \\
rosette~O & clumpy & -0.35 & 1.26: & 2.22: & 2.88: & 0.75 & 1.76 & 2.2 & 6.23 \\
lagoon~A & clumpy & -0.02 & 1.93 & 2.23 & 2.55 & 0.55 & 0.84 & 1.4 & 6.34 \\
lagoon~B & clumpy & -0.62 & 2.20 & 3.69 & 4.62 & 0.28 & 1.22 & 1.8 & 6.15 \\
lagoon~C & clumpy & -0.16 & 1.93 & 2.51 & 2.97 & 0.52 & 0.85 & 1.4 & 6.20 \\
lagoon~D & clumpy & -0.52 & 1.52 & 2.82 & 3.65 & 0.41 & 0.78 & 1.3 & 6.25 \\
lagoon~E & clumpy & 0.24 & 2.32 & 2.10 & 2.16 & 0.6 & 0.83 & 1.3 & 6.28 \\
lagoon~F & clumpy & 0.56 & 2.86 & 2.00 & 1.74 & 0.36 & 0.80 & 1.3 & 6.36 \\
lagoon~G & clumpy & -0.58 & 1.63 & 3.04 & 3.92 & 0.4 & 0.83 & 1.3 & 6.34 \\
lagoon~H & clumpy & -0.03 & 2.36 & 2.67 & 3.00 & 0.2 & 0.79 & 1.3 & 6.32 \\
lagoon~I & clumpy & 0.12 & 2.53 & 2.56 & 2.74 & 0.1 & 0.83 & 1.3 & 6.32 \\
lagoon~J & clumpy & -0.01 & 2.09 & 2.36 & 2.67 & 0.05 & 0.87 & 1.3 & 6.43 \\
lagoon~K & clumpy & 0.13 & 2.39 & 2.38 & 2.55 & 0.45 & 1.03 & 1.4 & 6.15 \\
ngc2362~A & simple & -0.18 & 1.83 & 2.44 & 2.93 & 0.38 & 0.65 & 1.2 & 6.50 \\
ngc2362~B & simple & 0.21 & 2.60 & 2.43 & 2.53 & 0.1 & 0.62 & 1.09 & 6.46 \\
dr21~A & chain & -0.35 & $>$1.32 & $>$2.27 & $>$2.92 & 0.69 & 1.97 & 2.3 & 5.78 \\
dr21~B & chain & -0.63 & $>$0.89 & $>$2.38 & $>$3.31 & 0.34 & 2.04 & 2.8 & \nodata \\
dr21~C & chain & -0.64 & 1.56: & 3.09: & 4.03: & 0.41 & 3.00 & 3.5 & \nodata \\
dr21~D & chain & -0.37 & 2.25 & 3.25 & 3.92 & 0.61 & 2.92 & 4.0 & 5.84 \\
dr21~E & chain & -0.18 & 2.30 & 2.92 & 3.40 & 0.44 & 2.36 & 3.7 & 6.00 \\
dr21~F & chain & -0.79 & $>$1.11 & $>$2.74 & $>$3.83 & 0.55 & 2.55 & 4.0 & \nodata \\
dr21~G & chain & -0.32 & $>$1.52 & $>$2.40 & $>$3.02 & 0.22 & 2.51 & 3.3 & \nodata \\
dr21~H & chain & -0.52 & 1.77: & 3.06: & 3.87: & 0.53 & 2.96 & 3.3 & \nodata \\
dr21~I & chain & -0.37 & $>$1.51 & $>$2.51 & $>$3.18 & 0.32 & 2.66 & 3.0 & 6.04 \\
rcw38~A & halo & 0.86 & 3.68 & 2.01 & 1.35 & 0.19 & 1.36 & 2.2 & 6.28 \\
rcw38~B & core & -0.53 & 3.51 & 4.61 & 5.34 & 0.36 & 1.06 & 2.6 & 6.28 \\
rcw38~C & clumpy & -0.22 & 2.21 & 2.68 & 3.09 & 0.78 & 1.62 & 2.5 & 6.28 \\
ngc6334~A & chain & -0.31 & 2.23 & 3.11 & 3.72 & 0.31 & 1.44 & 2.1 & \nodata \\
ngc6334~B & chain & 0.16 & 2.62 & 2.56 & 2.70 & 0.25 & 1.44 & 2.0 & 6.36 \\
ngc6334~C & chain & -0.64 & 2.07 & 3.61 & 4.56 & 0.15 & 1.10 & 1.8 & \nodata \\
ngc6334~D & chain & -0.51 & 2.08: & 3.35: & 4.17: & 0.09 & 1.88 & 2.8 & \nodata \\
ngc6334~E & chain & -0.03 & 2.74 & 3.05 & 3.37 & 0.24 & 1.88 & 3.1 & \nodata \\
ngc6334~F & chain & -0.09 & 2.18 & 2.61 & 3.00 & 0.17 & 1.28 & 1.8 & \nodata \\
ngc6334~G & chain & -0.37 & 2.41 & 3.4 & 4.07 & 0.26 & 1.61 & 2.5 & \nodata \\
ngc6334~H & chain & -0.22 & 2.34 & 3.03 & 3.55 & 0.26 & 1.38 & 1.8 & 6.20 \\
ngc6334~I & chain & -0.35 & 1.91: & 2.86: & 3.52: & 0.32 & 1.01 & 1.6 & \nodata \\
ngc6334~J & chain & 0.05 & 3.02 & 3.17 & 3.42 & 0.65 & 2.27 & 3.2 & 6.18 \\
ngc6334~K & chain & -0.34 & 1.66: & 2.60: & 3.24: & 0.31 & 3.00 & 3.1 & \nodata \\
ngc6334~L & chain & -0.00 & 2.51 & 2.77 & 3.07 & 0.16 & 2.44 & 3.2 & 5.84 \\
ngc6334~M & chain & -0.15 & $>$1.24 & $>$1.79 & $>$2.24 & 0.2 & 2.03 & 2.7 & \nodata \\
ngc6334~N & chain & 0.05 & $>$1.41 & $>$1.56 & $>$1.81 & 0.3 & 2.20 & 1.6 & \nodata \\
ngc6357~A & simple & -0.01 & 3.09 & 3.37 & 3.68 & 0.22 & 1.26 & 1.9 & 6.15 \\
ngc6357~B & simple & 0.27 & 2.96 & 2.67 & 2.70 & 0.36 & 1.30 & 1.9 & 6.15 \\
ngc6357~C & clumpy & 0.04 & 3.02 & 3.2 & 3.46 & 0.19 & 1.29 & 1.9 & 6.08 \\
ngc6357~D & clumpy & -0.72 & 2.42 & 4.13 & 5.16 & 0.38 & 1.27 & 2.0 & 6.04 \\
ngc6357~E & clumpy & 0.24 & 2.66$^\dagger$& 2.43$^\dagger$ & 2.49$^\dagger$ & 0.83 & 1.33 & 1.9 & 6.15 \\
ngc6357~F & simple & 0.02 & 3.19 & 3.40 & 3.68 & 0.5 & 1.41 & 2.1 & 6.18 \\
eagle~A & core & -0.33 & 2.24 & 3.14 & 3.77 & 0.01 & 0.88 & 1.4 & 6.38 \\
eagle~B & halo & 0.40 & 3.36 & 2.81 & 2.71 & 0.5 & 0.98 & 1.5 & 6.32 \\
eagle~C & clumpy & -0.19 & 2.07 & 2.71 & 3.21 & 0.6 & 0.96 & 1.4 & 6.23 \\
eagle~D & clumpy & 0.62 & 3.15 & 2.17 & 1.85 & 0.61 & 1.10 & 1.6 & 6.40 \\
eagle~E & clumpy & -0.53 & $>$1.53 & $>$2.84 & $>$3.67 & 0.07 & 2.40 & 2.6 & 6.00 \\
eagle~F & clumpy & 0.23 & 2.32 & 2.12 & 2.19 & 0.13 & 1.61 & 2.3 & \nodata \\
eagle~G & clumpy & -0.24 & 1.75: & 2.49: & 3.03: & 0.51 & 2.59 & 3.7 & \nodata \\
eagle~H & clumpy & -0.42 & $>$1.35 & $>$2.45 & $>$3.17 & 0.26 & 1.57 & 1.6 & \nodata \\
eagle~I & clumpy & 0.02 & 1.91 & 2.12 & 2.40 & 0.64 & 1.64 & 2.4 & 5.90 \\
eagle~J & clumpy & -0.04 & 1.59: & 1.92: & 2.25: & 0.41 & 1.51 & 2.1 & \nodata \\
eagle~K & clumpy & -0.61 & $>$1.35 & $>$2.82 & $>$3.73 & 0.04 & 1.42 & 2.0 & 6.28 \\
eagle~L & clumpy & -0.92 & $>$1.14 & $>$3.25 & $>$4.47 & 0.31 & 2.01 & 3.0 & \nodata \\
m17~A & clumpy & -0.57 & 2.20: & 3.60: & 4.47: & 0.03 & 1.92 & 3.2 & \nodata \\
m17~B & clumpy & -1.02 & 1.97: & 4.27: & 5.59: & 0.1 & 2.49 & 2.9 & \nodata \\
m17~C & clumpy & -0.26 & 2.62 & 3.39 & 3.95 & 0.09 & 1.70 & 2.5 & 6.15 \\
m17~D & clumpy & 0.05 & 3.19 & 3.33 & 3.58 & 0.07 & 1.53 & 2.4 & 6.04 \\
m17~E & clumpy & -0.49 & 2.23 & 3.47 & 4.27 & 0.18 & 1.56 & 2.5 & 6.38 \\
m17~F & clumpy & -0.74 & 2.12 & 3.86 & 4.90 & 0.03 & 1.70 & 4.4 & \nodata \\
m17~G & clumpy & -0.83 & 2.01 & 3.92 & 5.05 & 0.05 & 1.55 & 3.9 & \nodata \\
m17~H & clumpy & -0.35 & 2.62 & 3.58 & 4.23 & 0.07 & 1.25 & 2.2 & 6.00 \\
m17~I & clumpy & -0.17 & 2.72 & 3.32 & 3.80 & 0.02 & 1.45 & 2.1 & 6.15 \\
m17~J & clumpy & -1.37 & \nodata & \nodata & \nodata & 0.02 & 1.56 & 2.0 & \nodata \\
m17~K & clumpy & -0.16 & 2.94 & 3.52 & 3.98 & 0.14 & 1.41 & 2.2 & 6.00 \\
m17~L & clumpy & -0.26 & 3.13 & 3.91 & 4.47 & 0.05 & 1.82 & 2.8 & 6.08 \\
m17~M & clumpy & -0.15 & 2.66 & 3.21 & 3.66 & 0.21 & 1.83 & 2.5 & \nodata \\
m17~N & clumpy & -0.30 & 2.50 & 3.35 & 3.94 & 0.27 & 1.33 & 2.1 & 6.20 \\
m17~O & clumpy & -0.50 & 2.48 & 3.74 & 4.54 & 0.18 & 1.42 & 2.4 & 5.84 \\
carina~A & clumpy & -0.29 & 2.21 & 3.04 & 3.63 & 0.55 & 1.03 & 1.6 & 6.45 \\
carina~B & halo & 0.35 & 3.39 & 2.94 & 2.89 & 0.35 & 0.97 & 1.5 & 6.43 \\
carina~C & core & -0.10 & 3.13 & 3.58 & 3.98 & 0.17 & 0.94 & 1.4 & 6.18 \\
carina~D & chain & 0.23 & 2.71 & 2.50 & 2.58 & 0.5 & 0.90 & 1.4 & 6.38 \\
carina~E & chain & 0.05 & 2.53 & 2.70 & 2.95 & 0.05 & 0.84 & 1.4 & 6.38 \\
carina~F & chain & 0.31 & 2.61 & 2.25 & 2.24 & 0.36 & 0.94 & 1.5 & 6.58 \\
carina~G & chain & 1.24 & 3.94$^\dagger$ & 1.71$^\dagger$ & 0.77$^\dagger$ & 0.81 & 0.91 & 1.5 & 6.53 \\
carina~H & chain & 0.02 & 2.67 & 2.90 & 3.18 & 0.23 & 0.83 & 1.3 & 6.45 \\
carina~I & chain & -0.09 & 2.34 & 2.78 & 3.17 & 0.24 & 0.80 & 1.3 & 6.68 \\
carina~J & chain & 0.28 & 2.90 & 2.60 & 2.62 & 0.09 & 0.96 & 1.5 & 6.36 \\
carina~K & chain & 0.08 & 2.66 & 2.76 & 2.99 & 0.28 & 0.89 & 1.5 & 6.56 \\
carina~L & chain & 0.23 & 2.87 & 2.67 & 2.75 & 0.27 & 0.91 & 1.5 & 6.43 \\
carina~M & chain & 0.20 & 2.75 & 2.59 & 2.69 & 0.32 & 0.95 & 1.6 & 6.40 \\
carina~N & chain & 0.19 & 2.77 & 2.64 & 2.75 & 0.03 & 1.51 & 2.1 & \nodata \\
carina~O & chain & -0.15 & 2.62 & 3.17 & 3.62 & 0.09 & 1.15 & 1.8 & 6.04 \\
carina~P & chain & 0.52 & 3.11 & 2.33 & 2.10 & 0.5 & 0.87 & 1.5 & 6.62 \\
carina~Q & chain & 0.21 & 2.54 & 2.37 & 2.46 & 0.35 & 0.87 & 1.4 & 6.63 \\
carina~R & chain & 0.47 & 2.71 & 2.02 & 1.86 & 0.52 & 0.99 & 1.6 & 6.48 \\
carina~S & chain & 0.13 & 2.42 & 2.42 & 2.60 & 0.41 & 0.90 & 1.5 & 6.46 \\
carina~T & chain & 0.29 & 2.77 & 2.44 & 2.45 & 0.14 & 0.92 & 1.5 & 6.36 \\
trifid~A & clumpy & 0.52 & $>$2.04 & $>$1.26 & $>$1.04 & 0.81 & 1.53 & 1.3 & 6.28 \\
trifid~B & core & -0.29 & 2.48 & 3.32 & 3.91 & 0.2 & 0.92 & 1.4 & 6.28 \\
trifid~C & halo & 0.36 & 2.94 & 2.46 & 2.4 & 0.2 & 0.87 & 1.3 & 6.28 \\
trifid~D & simple & 0.29 & 2.44 & 2.10 & 2.11 & 0.68 & 1.27 & 1.4 & 6.28 \\
ngc1893~A & chain & 0.17 & 2.49 & 2.42 & 2.55 & 0.26 & 0.81 & 1.5 & 6.54 \\
ngc1893~B & chain & 0.11 & 2.70 & 2.74 & 2.93 & 0.21 & 0.81 & 1.5 & 6.42 \\
ngc1893~C & chain & -0.07 & 1.91: & 2.30: & 2.67: & 0.14 & 0.85 & 1.5 & 6.50 \\
ngc1893~D & chain & -0.98 & 1.34: & 3.55: & 4.83: & 0.09 & 0.75 & 1.4 & 6.28 \\
ngc1893~E & chain & -0.45 & $>$1.73 & $>$2.88 & $>$3.63 & 0.67 & 0.80 & 1.4 & \nodata \\
ngc1893~F & chain & -0.20 & 1.78: & 2.44: & 2.95: & 0.16 & 0.81 & 1.4 & 6.32 \\
ngc1893~G & chain & -0.03 & 2.29 & 2.61 & 2.95 & 0.33 & 0.89 & 1.6 & 6.18 \\
ngc1893~H & chain & 0.07 & 2.42 & 2.54 & 2.77 & 0.1 & 0.82 & 1.4 & 6.28 \\
ngc1893~I & chain & 0.20 & 2.77 & 2.63 & 2.74 & 0.33 & 0.83 & 1.5 & 6.45 \\
ngc1893~J & chain & -0.88 & 1.99 & 4.01 & 5.20 & 0.37 & 0.97 & 1.7 & 6.15 \\
\enddata
\tablecomments{
Properties of individual ellipsoidal subclusters in Paper~I. Column~1: Name of \mystix\ subcluster. Column~2: Fiducial subcluster radius ($=4r_c$). Column~3: Number of stars within 4 core radii. Column~4: Central surface density. Column~5: Estimated central volumetric density. Column~6: Ellipticity. Column~7: $J-H$ NIR absorption index. Column~8: X-ray median energy absorption indicator. Column~9: Median age.
}
\end{deluxetable}

\clearpage\clearpage

\section{Multivariate Analysis of Cluster Properties}

Figure~1---a multivariate ``pairs plot''---graphically displays the astrophysical properties from Table~1. The rows and columns in the $9\times9$ array of plots corresponds to individual variables, whose labels and units are shown in the panels on the diagonal. The panels on the diagonal shows single variable distributions as histograms, the lower triangle shows bivariate scatterplots with LOWESS non-parametric regression lines, and the upper triangle gives the statistical significance of the correlations using the Kendall's $\tau$ rank test. Graphs and computations were performed in the \Rlan\ statistical software environment (R Core Team 2014).

The LOWESS lines show the locally-weighted polynomial regression \citep{Becker88,Cleveland79,Cleveland81} for the variables of the rows (dependent variable) against the variables of the columns (independent variable). These lines are intended to guide the eye, rather than be used for statistical inference, and they may be inaccurate near the edges of the distribution where there are fewer points.

To test correlation between variables, our null hypothesis, $H_0$, is that the variables are uncorrelated, while our alternate hypothesis, $H_A$, is that a correlation exists, and $p(H_0)$ is the probability of obtaining a value of $\tau$ greater than or equal to the observed value. Null-hypothesis probabilities labeled $p\ge0.05$ do not pass the traditional threshold for rejection of $H_0$, probabilities with $0.05>p\ge0.0001$ are marginally statistically significant, while probabilities $p<0.0001$ indicate strong correlations. The \citet{Kendall38} rank test helps to reduce the effects of outlying points and uncertainties in measurement, but could still be vulnerable to correlated uncertainties. For example, the cluster radii and the central surface density (obtained from the models in Paper~I) are used to calculate the numbers of stars and the central volume density. On the other hand, some quantities like $ME$ and $J-H$ are obtained from independent astronomical measurements, so correlations between these quantities and other parameters are less likely to be a result of systematic biases. For the hypothesis test, values are rounded to one significant figure beyond the decimal point, representing the approximate precision of the data. The rounding does not affect the p-values in an appreciable way (i.e., few ties are produced in for the rank test).

\begin{figure}
\centering
\includegraphics[angle=90.,width=7.0in]{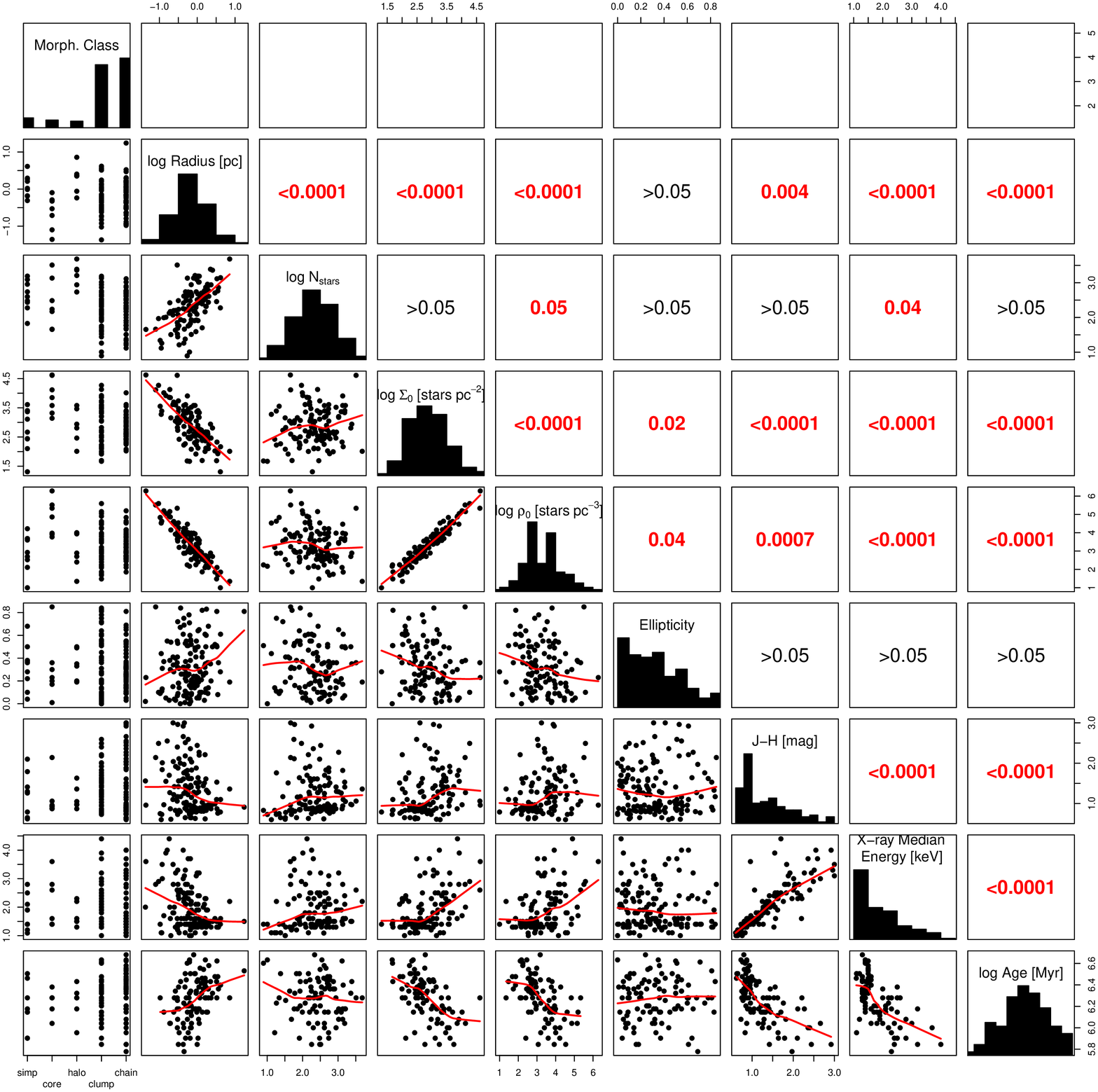}
\caption{Pairs
\label{pairs.fig}}
\end{figure}

Below, we summarize the statistical correlations with various variables that can be seen from Figure~1. 

\begin{description}
\item[Radius] There is a highly statistically significant positive correlation between subcluster radius and age, which could be explained astrophysically by subcluster expansion. There are also strong correlations with number of stars (positive), central surface density (negative), and central volume density (negative). Less obvious to the eye, but still statistically significant, are the negative correlations with the $J-H$ and $ME$ absorption indicators.
\item[Ellipticity] The ellipticities measured for subclusters show little correlation with any of the other subcluster properties. There is a marginal indication ($p<0.05$) that the denser subcluster are also more spherical, but there is much scatter in these relations.
\item[Number of Stars] There is a lot of scatter in many of the correlations between the number of stars in a subcluster and other subcluster properties, which is somewhat surprising given that this is one of the most fundamental subcluster properties. There is a marginally significant negative correlation between the number of stars and the central volume density of the cluster, which will be investigated in more depth in a later section.  
\item[Density] There is a tight correlation between the surface density and volume density, which is not a surprising result, but would not necessarily be true for every possible configuration of subcluster structure. There is also a statistically significant trend that the denser subclusters have higher absorption indicated by $J-H$ and $ME$.
\item[$J-H$ and $ME$] There is a tight correlation between these two indicators of absorption by the molecular cloud, $J-H$ tracing NIR dust absorption and $ME$ tracing X-ray gas absorption, with a few outlying points. The strong inverse relation between subcluster age and absorption was noted by \citet{Getman14}.
\item[Age] Subcluster age has statistically significant relations with subcluster radius (positive), surface density (negative), volume density (negative),  $J-H$ (negative), and $ME$ (negative); which indicates that subcluster age may be an important explanatory variable for subcluster properties. The age of a subcluster appears to have little relation to the number of stars in the subcluster.
\item[Morphological class] These classes are categorical labels without an intrinsic ordering (1=``simple,'' 2=``core,'' 3=``halo,'' 4=``clumpy,'' 5=``chain''), so we do not compute $p$-values using Kendall's $\tau$. Regarding radius, ``simple'' subclusters are similar in size to the ``halo'' component of core-halo structures, while the ``core'' components are a factor of $\sim$10 smaller. The ``clumpy'' and ``chain'' subclusters cover this whole range in size. This trend is again reflected in the central surface and volume densities. Regarding the absorption indicators, $J-H$ and $ME$, ``simple'' and ``core-halo'' subclusters typically have lower absorptions, while ``clumpy'' and ``chain'' subclusters have a wider range of absorptions. Regarding age, there is no obvious difference in typical age for the different groups; however, the ``clumpy'' and ``chain'' subclusters have a wider range of ages.
\end{description}

\subsection{Principal Component Analysis}

Principal component analysis (PCA) is performed using the variables $\log r_4$, $\log n_4$, $\epsilon$, $ME$ and $\log \mathrm{Age}$ (we exclued $\log \Sigma_0$, $\log \rho_0$, and $J-H$ because these quantities are redundant). Subclusters with missing data are excluded from this analysis. {The distribution of $ME$ deviates strongly from normality; most subclusters are lightly obscured. $ME$ is peaked at a low value of 1.4~keV, but there is a tail of highly obscured subclusters out to 3.0 keV. Instead we use a standardized variable based on the rank of $ME$ values, which is normally distributed with a standard deviation of 0.5 (similar to other variables) as provided by the R-code below.
\begin{verbatim}
ME.standardized <- qnorm((rank(ME)-0.5)/length(ME))/0.5
\end{verbatim}
Table~\ref{pca.tab} provides PCA loadings for the first three components. The first two principal components, $Comp.1$ and $Comp.2$, account collectively for 84\% of the variance, shown in the bar-chart in Figure~\ref{pca.fig} (left), so the remaining components are not particularly important. The components $Comp.1$ and $Comp.2$ are both linear combinations of radius, number of stars, median energy and age, while the third principal component, $Comp.3$, also includes ellipticity.\footnote{When principal component analysis is performed using the original $ME$ rather than the standardized $ME$, the relative relations between the variables remain the same albeit rotated in the ($Comp.1$,~$Comp.2$) plane.}  This demonstrates that the global distribution of subcluster properties can be reduced down to two variables, the first one being a combination of age, absorption, and central (surface) density, and the second one being the number of stars in a cluster.

Figure~\ref{pca.fig} (right) is the biplot for the first two components, $Comp.1$ and $Comp.2$); i.e., it shows the cluster parameters projected into the ($Comp.1$,~$Comp.2$) plane using the biplot definition from \citet{Gabriel71}. Red arrows show the original variables projected in this coordinate system. The arrows for $\log \mathrm{Age}$, $ME$, and $\log \Sigma_0$ are almost parallel ($\log \rho_0$ points more-or-less in the same direction), which indicates that subcluster age, density, and absorption are all closely related. In contrast, the $\log n_4$ arrow is nearly orthogonal, indicating that---globally---the number of stars in a cluster is not strongly affected by subcluster age, surface density, or absorption. The $r_4$ arrow lies in between these two axes, reflecting our earlier finding of a statistically significant relations between age and radius (positive) and radius and number of stars (positive).

\begin{deluxetable}{lrrr}
\tablecaption{PCA of Subcluster Properties \label{pca.tab}}
\tabletypesize{\small}\tablewidth{0pt}
\tablehead{
\colhead{Property} &  \colhead{Comp.1} &  \colhead{Comp.2} &  \colhead{Comp.3}\\
\colhead{Variance} &  \colhead{56\%} &  \colhead{28\%} &  \colhead{9\%}
}
\startdata
$\log r_4$ &			$-$0.361		& $-$0.471	& $-$0.630\\
$\log n_4$ &			$-$0.892		& \nodata	& 0.350\\
$\epsilon$ &			\nodata	& \nodata		& -0.549\\
$ME_\mathrm{st.}$ &				$-$0.270		& 0.812		& -0.384\\
$\log \mathrm{Age}$ &	\nodata	& $-$0.336	& $-$0.179\\
\enddata
\end{deluxetable}

\begin{figure}
\centering
\includegraphics[angle=0.,width=3.0in]{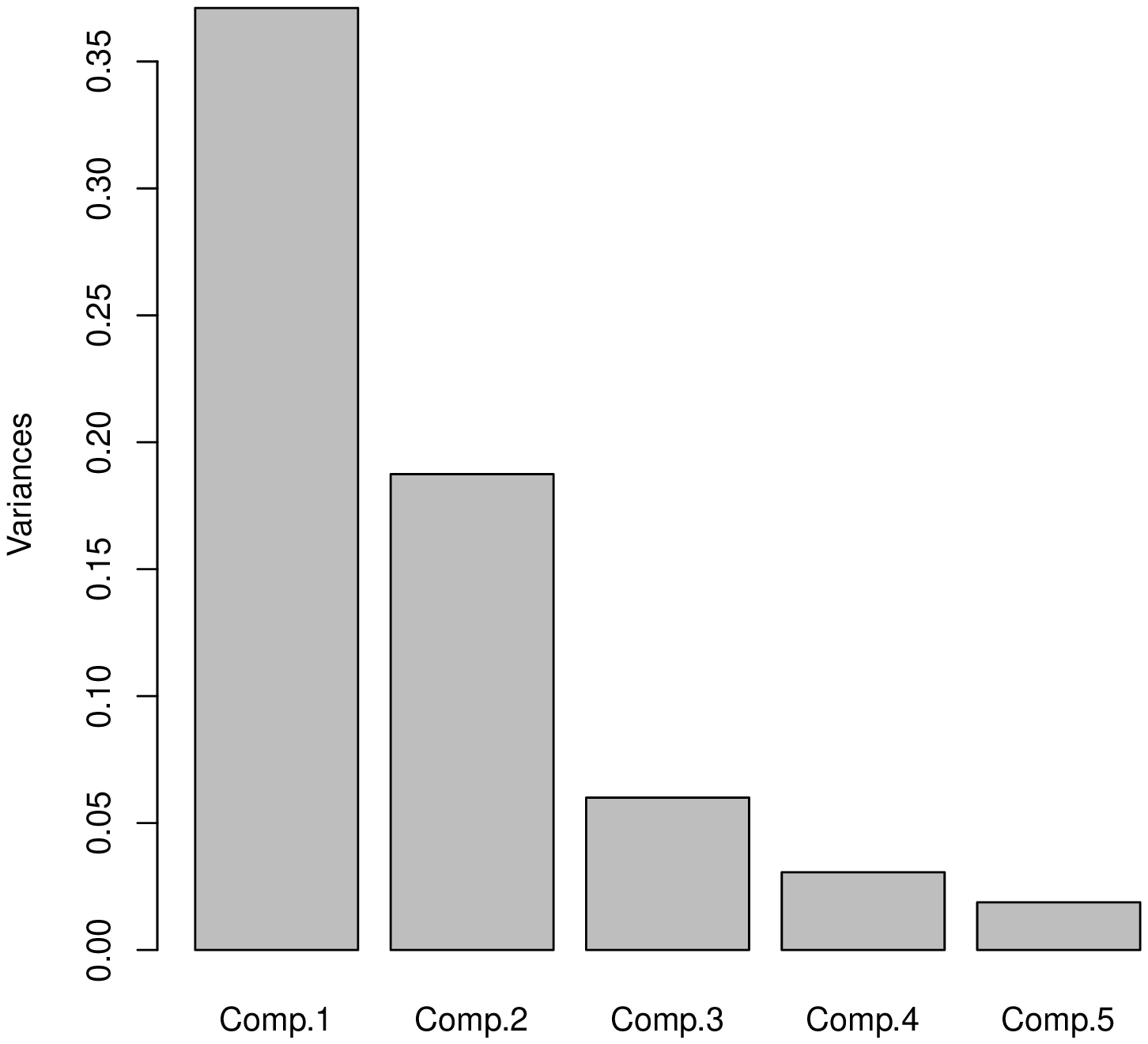}
\includegraphics[angle=0.,width=3.0in]{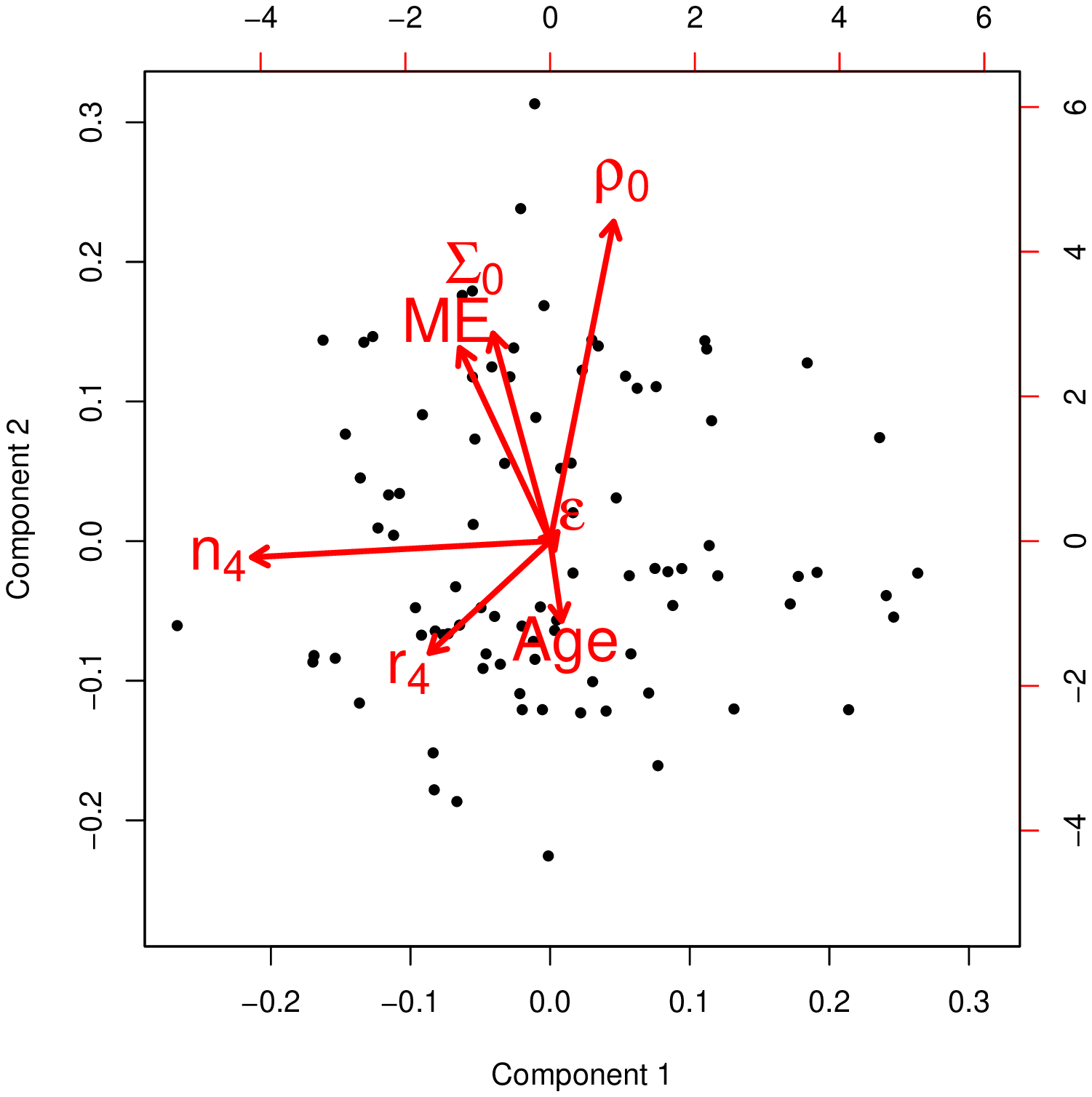}
\caption{Left: A scree plot showing the relative contributions of the five PCA components to the total variance. Right: A biplot shows the subcluster observations (black points) and the Table~1 variables (red arrows) projected onto the plane of the first two PCA components. The variables for age, absorption, and stellar density are roughly parallel, while the number of stars in a cluster is orthogonal, and cluster radius is in between.
\label{pca.fig}}
\end{figure}

\clearpage\clearpage

\subsection{The Subcluster Age--Radius Relation}

Figure~\ref{age_radius.fig} shows the relation between $\log r_4$ and $\log \mathrm{Age}$. Although there is significant scatter in this relation, there is a clear, positive trend between subcluster median age from the $AgeJX$ analysis and subcluster size. This can be interpreted as cluster expansion---a phenomenon expected to occur as gas is removed from young embedded clusters \citep[e.g.][]{Tutukov78,Hills80,Goodwin06,Baumgardt07,Moeckel10,Bastian11}. Scatter in the relation is caused by uncertainty in estimating stellar ages and subcluster radii, but it can also be intrinsic; i.e.\ subclusters form with different initial radii, or expansion happens at different rates for different subclusters. We fit a simplistic linear regression model, which helps provide intuition about this relation, even if it is not physically realistic. For this regression analysis, both variables have uncertainty and it is unclear if one variable can be labeled the ``independent'' variable and the other the ``dependent'' variable; thus, we show the results of several linear regression strategies from \citet{Isobe90}. 

The gray, dashed lines on the plot indicate the size of a sphere expanding from the center of the cluster at a constant velocity of 0.5, 1, 2, and 4~km~s$^{-1}$. Most of the regression lines are slightly steeper than these lines (the ordinary least squares regression does not account for uncertainty in median age, so its slope is likely to not be steep enough). Few points lie above the 1~km~s$^{-1}$ line on the left side of the plot, but 10 lie in this region on the right side of the plot. However, the OLS bisector line and the reduced major-axis line, which look like reasonable fits, have slopes, $\beta$, close to the $\beta=1$ constant velocity slope. The parameters for the reduced-major axis regression line are provided in Table~\ref{regressions.tab}.

A simple model for subcluster expansion is the case where the kinetic energy of the stars is much greater than the potential energy of the clusters, so the stars coast outward at a roughly constant velocity. Many studies of stellar velocities in young stellar clusters, both observational \citep[Orion and NGC2264;][]{Furesz06,Furesz08,Tobin15} or theoretical \citep[e.g.,][]{Bate03,Bate09a}, show initial stellar velocities of $>$3~km~s$^{-1}$. However, some young stellar clusters have lower measured velocity dispersions, like $\gamma$~Velorum with 0.34 and 1.60~km~s$^{-1}$ components \citep{Jeffries14}. Almost all of the MYStIX subclusters on the plot lie below the 1~km~s$^{-1}$ line. If a cluster were unbound with stars moving at velocities $>$3~km~s$^{-1}$, it is unlikely that so many stars (the $r_4$ radius contains on the order of half the stars in the stars in the cluster) would be so near the location at which they were formed. This suggests that the stars are not just freely drifting, but it is likely that most of the subclusters are currently gravitationally bound.\footnote{Alternatively, some of the subclusters that have small $r_4$ for their $Age_{JX}$ values, could have recently become unbound. Nevertheless, the unbinding event (by gas expulsion) would have had to happen very recently compared to the subclusters' total ages for this trend to still be seen. It is unlikely that the majority of subclusters happen to be seen in a state immediately after gas expulsion.} Confirmation of the bound or unbound state of these clusters would require radial velocity observations.

The expansion rate of bound young stellar clusters was also investigated by \citet{Moeckel10}, who take the cluster of \citet{Bate09a} and evolve it forward
in time to 10~Myr under a variety of assumptions about gas removal timescales.
They find (e.g., their Figure~5) that although the initial velocity dispersion is several
km~s$^{-1}$, the velocity dispersion decreases as the cluster expands to around 1~km~s$^{-1}$ after 2~Myr.  This is reflected in the size of the clusters with time (e.g., their Figures~4 and 6) where the half-mass radius typically expands by a factor of 10 between
1 and 10 Myr (typically at a rate of $\sim$1 km~s$^{-1}$).  These expansion rates are 
consistent with our observations. 

The age--radius scatter plot shows little evidence for larger radii with age until 2~Myr ($\log\mathrm{Age}=6.3$), which is also consistent with the simulation results. The LOWESS curve for this relation (Figure~\ref{pairs.fig}) also shows a hint of concavity at $\sim$2~Myr.

\begin{deluxetable}{lrrr}
\tablecaption{Relations of of Subcluster Properties\label{regressions.tab}}
\tabletypesize{\small}\tablewidth{0pt}
\tablehead{
\colhead{Relation} &  \multicolumn{2}{c}{Parameters} &  \colhead{Stat.}\\
\cline{2-3}
\colhead{} &  \colhead{$\log A$} &  \colhead{$\alpha$} &  \colhead{Signif.}\\
\colhead{(1)} &  \colhead{(2)} &  \colhead{(3)} &  \colhead{(4)}
}
\startdata
$\log r_4=\log A + \alpha \log \mathrm{Age}$ 	&	$-18.0\pm0.1$	& $2.9\pm0.4$	& $<$0.0001	\\
$\log \Sigma_0=\log A+ \alpha \log r_4$ 		&	$2.6\pm0.1$	& $-1.9\pm0.2$	& $<$0.0001	\\
$\log \rho_0=\log A+ \alpha \log r_4$ 		&	$2.9\pm0.2$	& $-2.6\pm0.1$	& $<$0.0001	\\
$\log n_4=\log A+\alpha \log r_4$ 			&	$2.5\pm0.1$	& $1.7\pm0.3$	& $<$0.0001	 \\
\enddata
\tablecomments{Reduced-major axis regression lines. Column~1: Formula for the relation between the two quantities. Column~2: Scale. Column~3: Power-law index and uncertainty. Column~4: The null-hypothesis probability, $p(H_0)$, that a Kendall's $\tau$ statistic greater or equal to the calculated value would be produced.}
\end{deluxetable}
\clearpage\clearpage

\begin{figure}
\centering
\includegraphics[angle=0.,width=3.0in]{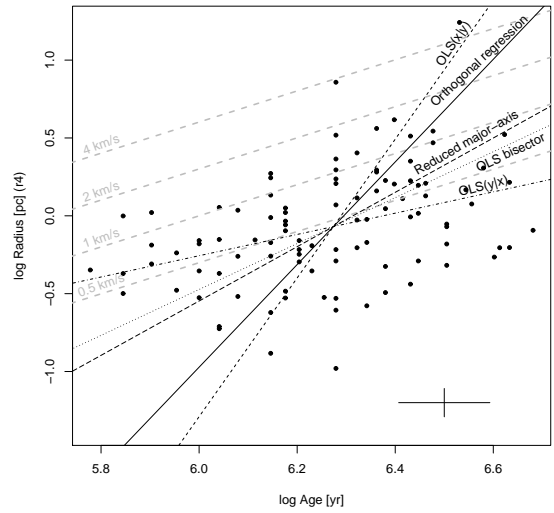}
\caption{Scatter plot of subcluster median age vs.\ radius, $r_4$. Linear regression lines are shown as black lines: OLS$(y|x)$ (dot-dashed), OLS$(x|y)$ (dashed), OLS bisector (dotted), orthogonal regression (solid), reduced major-axis (long dashed). Gray dashed lines indicate the age--radius relation for a sphere expanding from the origin at a constant velocity of 0.5, 1, 2, and 4~km~s$^{-1}$.\label{age_radius.fig}}
\end{figure}

\clearpage\clearpage

\subsection{The Radius--Central Density Relation}

If clusters are expanding, the density of stars in the center of the cluster will become attenuated with increasing radius. Figure~\ref{radius_density.fig} shows the surface density, $\Sigma_0$, radius relation and the volume density, $\rho_0$, radius relation for subclusters. Both these graphs contain the same information, since volume density is inferred from surface density and radius. However, volume density is more astrophysically meaningful, while surface density can be estimated without assuming a subcluster radius. The distribution of points suggests that the relation between radius and density is nearly a power law. The same linear regressions strategies that were used above are used here (the black lines on the plot).

If a subcluster were expanding isomorphically, with no stars gained or lost, the central surface density would decrease proportionally to $r^{-2}$ and the central volume density would decrease proportional to $r^{-3}$ (these lines are shown in gray). However, the observed relations are somewhat less steep, with $\Sigma_0\propto r^{-1.8}$ and $\rho_0 \propto r^{-2.3}$. \citet{Pfalzner11} notes a similar trend for young embedded clusters from \citep{Lada03} where $\rho_0 \propto r^{-1.3\pm0.7}$. 

These flatter slopes indicate that larger clusters contain more stars. The number of stars within a projected radius is a function of radius and surface density, so the $n_4\sim r_4$ relation shown in Figure~\ref{N_r.fig} is another method of displaying the information from Figure~\ref{radius_density.fig}. The parameters for the reduced-major axis regression lines described above are provided in Table~\ref{regressions.tab}.

\begin{figure}
\centering
\includegraphics[angle=0.,width=5.5in]{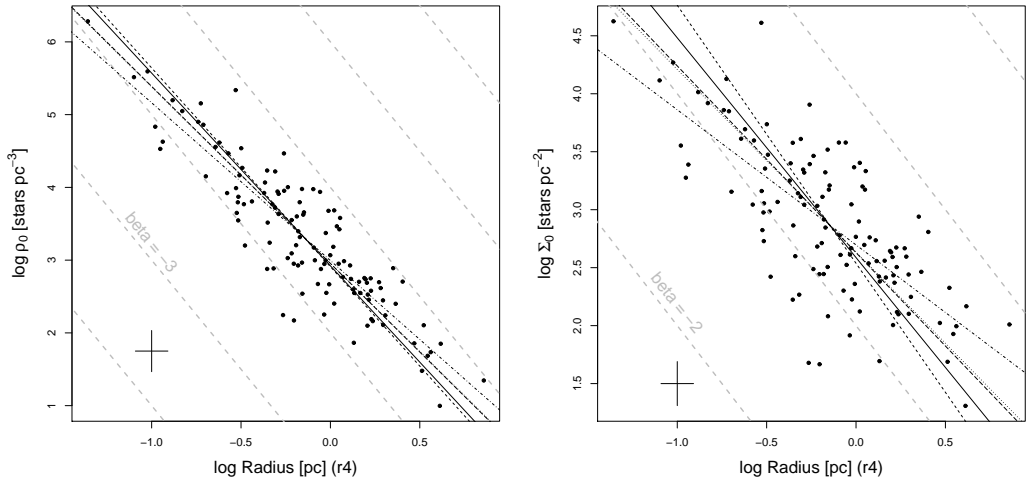}
\caption{Left: Scatterplot of central surface density, $\Sigma_0$, vs.\ subcluster radius, $r_4$. Right: Scatterplot of central volume density, $\rho_0$, vs.\ subcluster radius, $r_4$. Black lines indicate linear regression fits (line styles have the same meaning as in Figure~\ref{age_radius.fig}). Gray lines indicate density--radius tracks for an expanding cluster with a constant number of stars. \label{radius_density.fig}}
\end{figure}

\begin{figure}
\centering
\includegraphics[angle=0.,width=3in]{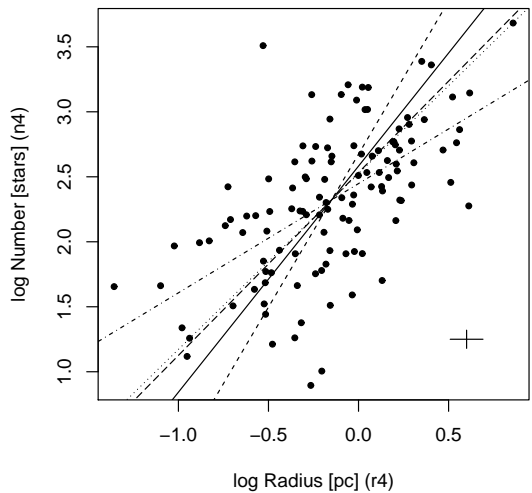}
\caption{Scatterplot of number of stars, $n_4$, vs.\ radius, $r_4$. Black lines indicate linear regression fits (line styles have the same meaning as in Figure~\ref{age_radius.fig}). \label{N_r.fig}}
\end{figure}

\clearpage\clearpage

\subsection{The Central Density--Number of Stars Relation}

From Figure~\ref{pairs.fig}, one can see that surface density $\Sigma_0$ is mostly independent of the number of stars in a cluster $n_4$, while there may be a slight negative relation between $\rho_0$ and $n_4$. To investigate this further, we partition the sample into a group with $n_4<200$~stars (54 subclusters) and a group with $n_4>200$~stars (63 subclusters). Figure \ref{N_density.fig} shows the $\rho_0\sim n_4$ relation, and the dashed horizontal lines indicate the median $\rho_0$ values for the two subsamples, revealing that the group of subclusters with more stars tend to have lower volume densities. 

We can test the statistical significance of this trend using the Anderson-Darling two-sample test---in Figure~\ref{rho0_adtest.fig} (left) the cumulative distributions of $\Sigma_0$ of the two samples ($n_4<200$~stars in black; $n_4>200$~stars in gray) are compared, showing very similar distributions with a $p$-value of 0.67 (not significant). In contrast, for the cumulative distributions of $\rho_0$, there is a visible shift, and the Anderson-Darling test gives a $p$-value of 0.01 (marginally significant).

The peak surface density can be measured independently from subcluster radius---i.e.\ Paper~I (their Figures~4 and 6) finds that there is good agreement between the central surface densities obtained by model fitting and non-parametric smoothing---so the $\Sigma_0\sim n_4$ relation may be more robust than the $\rho_0\sim n_4$ relation. 
If we consider $\Sigma_0$, $\rho_4$, and $r_4$ as functions of $n_4$, then we have $\Sigma_0(n_4) = 2 r_4(n_4)\rho_0(n_4)$. From Figure~\ref{N_r.fig} we see that that $r_4$ is an increasing function of $n_4$, plus scatter, which implies that $\rho_0$ must have a negative relation with $n_4$ in order to produce the observed flat $\Sigma_0\sim n_4$ relation. 


An inverse trend between volume density and number of stars may seem counterintuitive. Nevertheless, there are well-known examples that do exhibit the trend that we observe: in the Orion region, the Orion Hot Core surrounding the BN/KL object is denser than the much richer Trapezium cluster \citep{Rivilla13}. Note that the BN/KL object is likely behind the ONC cluster, and only coincidentally superimposed due to the direction that we are observing from. 

If the $\rho_0\sim n_4$ relation is valid, it could be explained by the merging of subclusters. 
Consider, for example, two subclusters with the same initial radii $r_\mathrm{init}$ and masses $M$. If they coalesce, but without time to dynamically relax, the initial potential energy will be approximately the same as the final potential energy due to the conservation of energy (i.e., the kinetic energies of the stars will only be slightly perturbed if the merger happens quickly enough). Thus, the initial potential energy when the clusters are far away from each other will be 
\begin{equation}
PE = kGM^2/r_\mathrm{init} + kGM^2/r_\mathrm{init} = 2kGM^2/r_\mathrm{init},
\end{equation}
where $G$ is the gravitational constant and $k$ is a constant of order unity. The final potential energy will be 
\begin{equation}
PE = kG(2M)^2/r_\mathrm{final}.
\end{equation}
Thus, $r_\mathrm{final} = 2\,r_\mathrm{init}$, and the ratios of final to initial density will be 
\begin{equation}\label{rho_evol.eqn}
\rho_\mathrm{final}/\rho_\mathrm{init} = (2M/M)(r_\mathrm{init}/r_\mathrm{final})^3 = 1/4.
\end{equation}
A similar effect is observed by \citet{Smith11} in simulations of interacting and merging stellar clusters. The encounters between subclusters create low-density halos of stars around the clusters which help facilitate the cluster mergers.

\begin{figure}
\centering
\includegraphics[angle=0.,width=3in]{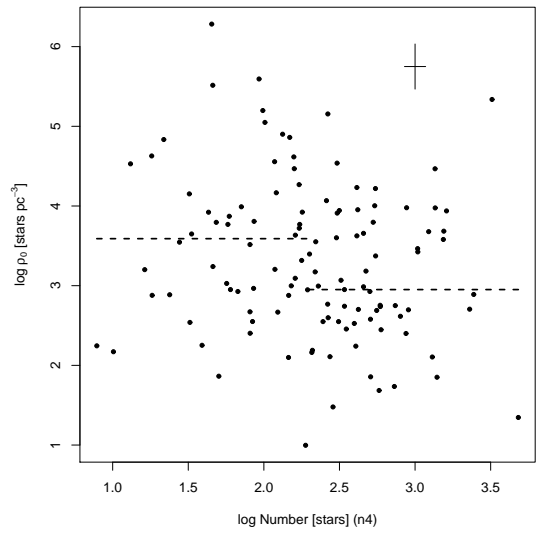}
\caption{Scatterplot of central volume density vs.\ the number of stars in a subcluster. The dashed lines indicate the median $\rho_0$ for subclusters with $<$200 stars and subclusters with $>$200 stars. \label{N_density.fig}}
\end{figure}

\begin{figure}
\centering
\includegraphics[angle=0.,width=6in]{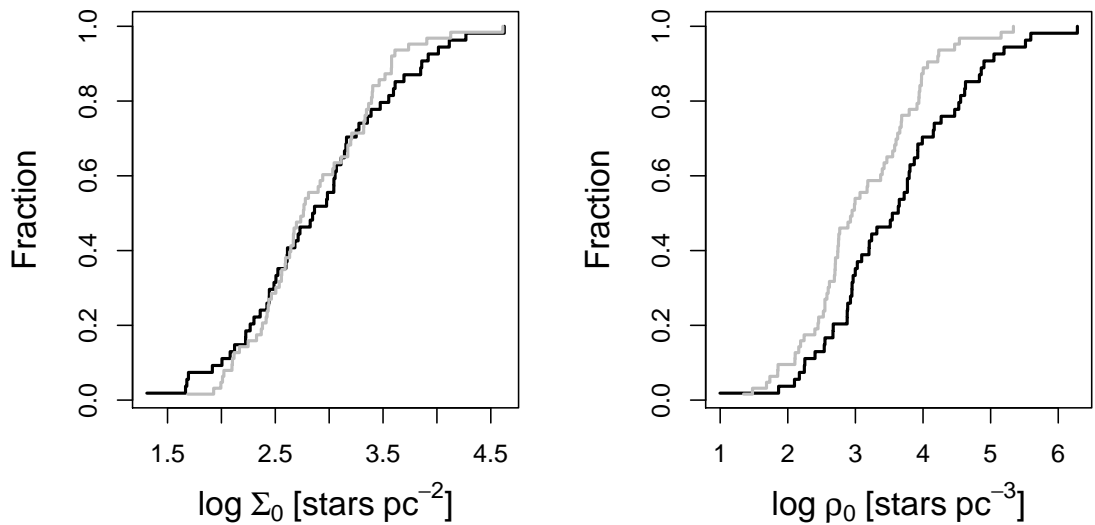}
\caption{Left: Cumulative distributions of $\Sigma_0$ for subclusters with $<$200 stars (black) and $>$200 stars (gray). The $p$-value for the two-sample Anderson--Darling test with the null hypothesis that the distributions are the same is $p=0.67$. Right: Cumulative distribution of $\rho_0$ for subclusters with $<$200 stars (black) and $>$200 stars (gray). The $p$-value for the two-sample Anderson--Darling test is $p=0.01$.\label{rho0_adtest.fig}}
\end{figure}

\clearpage\clearpage

\subsection{The Age--Number of Stars Relation}

When considering only the properties $n_4$ and $\mathrm{age}$ for all subclusters, as shown in the scatter plot in Figure~\ref{n_age.fig}, there is no sign of correlation between subcluster age and the number of stars in a subcluster. In addition, the principal component analysis in Figure~\ref{pca.fig} shows that age and number of stars are almost orthogonal quantities. Therefore, almost any combination of age and number of stars in a subcluster are possible. This is despite of the strong $\mathrm{age}\sim r_4$ relation and the strong $n_4 \sim r_4$ relation. 

This lack of correlation may appear to contradict the hierarchal mergers of subclusters scenario that was indicated by the subcluster ellipticity distribution (Paper~I), the $\rho_0\sim n_4$ relation, and the $n_4\sim r_4$ relation. 
Nevertheless, this result could still be consistent with a scenario in which individual subclusters grow in $n_4$ over time but collections of subclusters from different star-forming regions with different environments do not show a trend between age and $n_4$. For example, gas expulsion in some regions may slow their clusters' mergers \citep[e.g.,][]{Kruijssen12}, so different star forming regions at the same age could be at different points in their subclusters' growth. \citet{Fellhauer09} suggests that subcluster mergers happen quickly before gas expulsion but are impeded after gas expulsion, so subclusters may behave differently at the same age depending on when their gas was expelled. A possible example of these phenomena in \mystix\ can be seen in NGC~1893---this is one of the oldest star-forming regions, but it is still highly substructured, possibly due to early gas expulsion which has inhibited dynamical evolution of the stellar population. If we examine where subclusters from different MSFRs lie on the diagram, we find that they do not share the same locus. For example, the NGC~6357 subclusters all lie on the upper left side of the diagram, while the Rosette subclusters lie on the lower right. Thus, we cannot rule out subcluster mergers.

When we attempt to control for this effect by considering only subclusters that are highly embedded in molecular clouds (i.e., $ME>2.5$~keV) there is a marginally significant positive $ n_4 \sim \mathrm{age}$ relation ($p<0.05$), although this is not seen in the sample of subclusters that are not embedded. This hints that growth of subclusters in time does occur---for the embedded subcluster population---even if the effect does not appear as a global trend when comparing different MSFRs.

\begin{figure}
\centering
\includegraphics[angle=0.,width=3in]{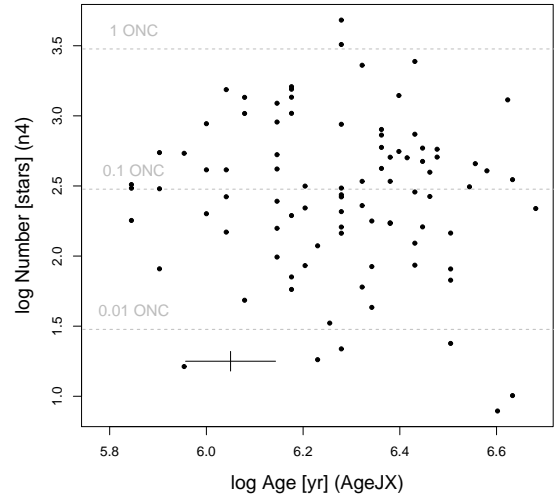}
\caption{Scatterplot of number of stars ($n_4$) vs.\ subcluster age. The gray dashed lines show tracks of constant population for subcluster's containing 100\%, 10\%, and 1\% of number of stars in the ONC ($\sim$3000~stars).  \label{n_age.fig}}
\end{figure}

\clearpage\clearpage

\subsection{The Effect of Morphological Classes on Relations of Physical Properties}

We consider the possibility star-forming regions with a different morphology of cluster structure (the ``morphological class'' property in Table~\ref{intrinsic.tab}) could be undergoing different types of dynamical evolution. Differences in young stellar cluster evolution have been reported in the literature, for example the three distinct young-stellar-cluster evolutionary tracks identified by \citet{Pfalzner11} and \citet{Pfalzner12} for embedded, starburst and non-starburst clusters. 
However, our ability to investigate the effect of morphological class on other subcluster properties is limited by our small samples of  ``simple'' and ``core-halo'' clusters. A single region with a ``chain'' or ``clumpy'' structure may have tens of subclusters, so their subclusters make up the majority objects of the catalog of 142 subclusters. On the other hand, as a result of small sample sizes, statistical tests using the  9 ``simple'' subclusters, 7 ``core'' subclusters, or 6 ``halo'' subclusters are likely to be inconclusive.

In Figure~\ref{morph1.fig} radius--density scatterplots are shown for four groups of subclusters, stratified by the different morphological classes. All regions reveal similar negative correlations between these two quantities, which are all statistically significant. However, there is some difference between the slopes of power-law regressions to these data, which are provided in Table~\ref{morph.tab}; confidence intervals (65\%) on the power-law indices are calculated from 1000 bootstrap iterations. The most prominent difference between the correlations is that the ``chain'' subclusters produce a less steep slope, more similar to what was found by \citet{Pfalzner11} for young embedded subclusters in the \citet{Lada03} sample. This indicates that subclusters in these star-forming regions, including NGC~2264, DR~21, NGC~6334, NGC~1893, and parts of Carina, are gaining more stars as they expand. In contrast, the subclusters in ``clumpy'' regions have a steeper relation which is consistent with expansion without gaining or losing stars. There are too few subclusters from the ``simple'' and ``core-halo'' categories to definitively determine which of the relations they more closely resemble---however the slopes of the regression-line fits more closely resemble the ``clumpy'' case. 

The data points in Figure~\ref{morph1.fig} are marked by cluster age, using smaller circles for younger subclusters and larger circles for older subclusters; subclusters without reliable age measurements are marked with ($+$) symbols. In the panel showing ``chain'' subclusters, there is a clear age progression, with younger subclusters in the upper left of the panel, and older subclusters in the lower right---which is what would be expected from a simple scenario of subcluster expansion. This trend is less clear in the other cases, and there are hints that it may even be reversed for the ``clumpy'' subclusters.

Figure~\ref{morph2.fig} shows the density--number-of-stars scatterplots for the different morphological classes. The negative trend is clearly visible for the ``chain'' subclusters at high statistical significance (Table~\ref{morph.tab}). The power-law index for this trend is $\alpha\approx-2$, which is close to the value obtained in Equation~\ref{rho_evol.eqn} from our hypothetical merger scenario. In the ``simple'' and ``core-halo'' cases, subclusters lie, more-or-less, within the same locus in this parameter space as the ``chain'' subclusters; however, there are too few data points to identify any trend. The subclusters in the ``clumpy'' sample are more spread out in this parameter space, and there is even the suggestion of the opposite trend in the opposite direction (not statistically significant).  

The subclusters on the Figure-\ref{morph2.fig} plots are also marked by age, as above. On the panel showing ``chain'' subclusters, the younger stars are located in the upper left and the older subclusters are located in the in the lower right, which is, again, the expected results from the picture of subcluster expansion and hierarchical mergers that we have been developing in this paper. This age trend is not clearly apparent in the other panels.

Paper~I suggests that the ``clumpy'' subclusters may be a different type of structure than the other subclusters. These subclusters correspond to modes in stellar surface-density maps that were judged to be real by our AIC analysis, but they are, nevertheless, often part of large-scale clusters of stars, as is the case for many of the ``clumpy'' subclusters in M~17. Thus, it may be incorrect to directly compare their properties to properties of ``simple,'' ``core-halo,'' or ``chain'' subclusters, which are usually discrete structures well separated from each other. This difference may affect how the ``clumpy'' class of subclusters behave.

\begin{deluxetable}{lrlr}
\tablecaption{Relations of Morphologically Stratified Subcluster Samples\label{morph.tab}}
\tabletypesize{\small}\tablewidth{0pt}
\tablehead{
\colhead{Morph.} &  \multicolumn{2}{c}{Parameters} &  \colhead{Stat.}\\
\cline{2-3}
\colhead{Class} &  \colhead{$\log A$} &  \colhead{$\alpha$} &  \colhead{Signif.}\\
\colhead{(1)} &  \colhead{(2)} &  \colhead{(3)} &  \colhead{(4)}
}
\startdata
\multicolumn{4}{l}{Density vs.\ Radius: $\log \rho_0 = \log A+ \alpha \log r_4$}\\
simple &	$3.2\pm1.1$	& $-3.7$:	& 0.01\\
core &	$3.4\pm0.6$	& $-2.2\pm0.7$	& 0.05 \\
halo &	$3.6\pm0.9$	& $-2.6$:	& 0.02\\
clumpy &	$2.7\pm0.5$	& $-3.1\pm0.3$	& $<$0.0001\\
chain &	$2.9\pm0.2$	& $-2.0\pm0.1$	& $<$0.0001\\
\cline{1-4}
\multicolumn{4}{l}{Density vs.\ Number: $\log \rho_0 = \log A+ \alpha \log n_4$}\\
chain &	$7.3\pm0.4$	& $-1.8\pm0.4$	& $<$0.0001\\
\enddata
\tablecomments{Reduced major-axis regression lines for subcluster samples stratified by the four morphological classes. Column~1: Morphological class. Column~2: Scale. Column~3: Power-law index and uncertainty. Column~4: The null-hypothesis probability, $p(H_0)$, from Kendall's $\tau$ test.}
\end{deluxetable}

\clearpage\clearpage

\begin{figure}
\centering
\includegraphics[angle=0.,width=6.0in]{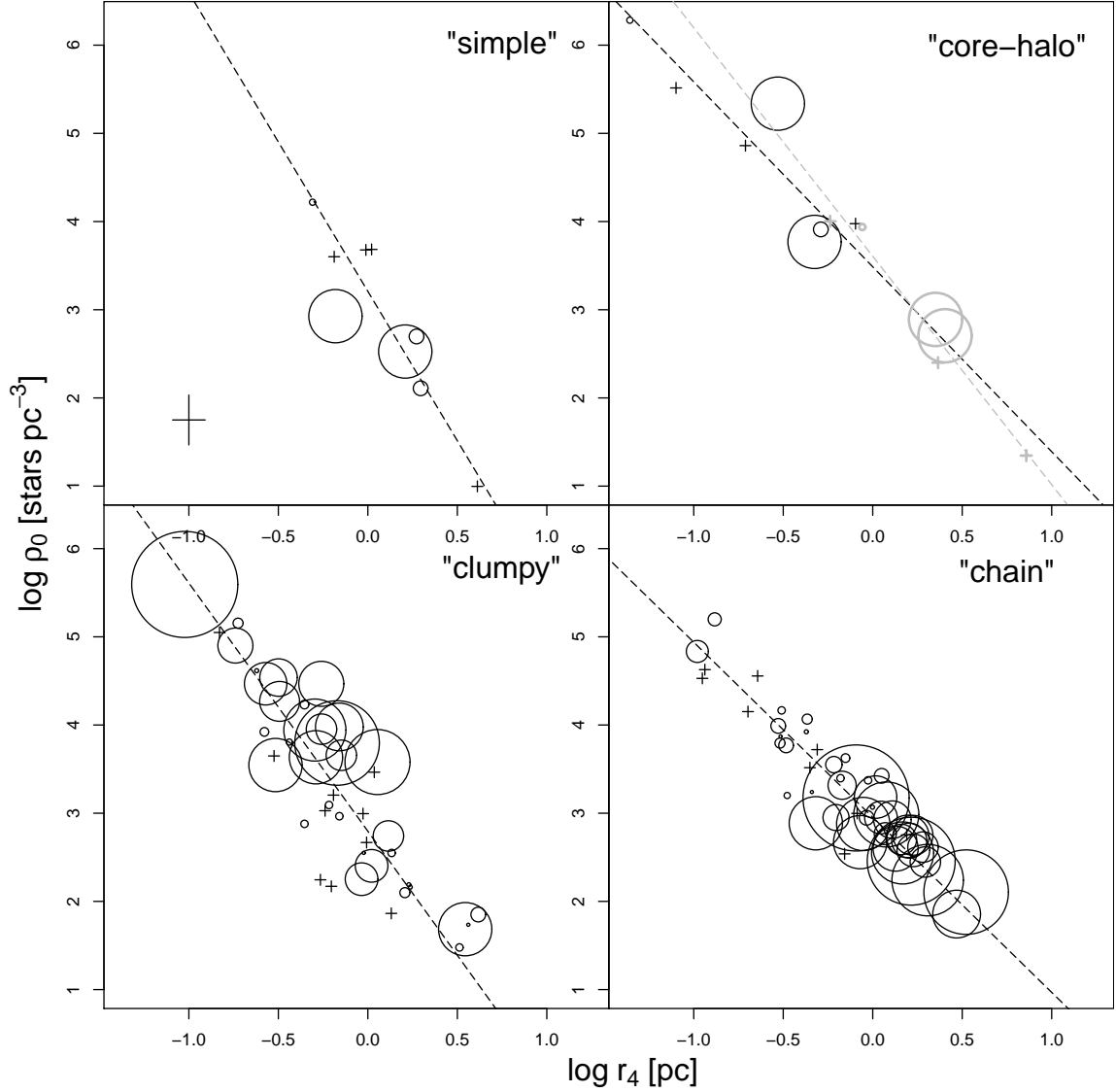}
\caption{Scatter plots of $\rho_0$ vs.\ $r_4$ stratified by the four classes---from left to right and top to bottom: ``simple,'' ``core-halo'' (black indicates ``core''; gray indicates ``halo''), ``clumpy,'' and ``chain.'' Symbol size indicates subcluster age, with smaller circles indicating younger subclusters, larger circles indicating older subclusters, and $+$'s marking subclusters with missing ages. The dashed lines show the reduced major axis regression line for statistically significant relations.
\label{morph1.fig}}
\end{figure}

\begin{figure}
\centering
\includegraphics[angle=0.,width=6.0in]{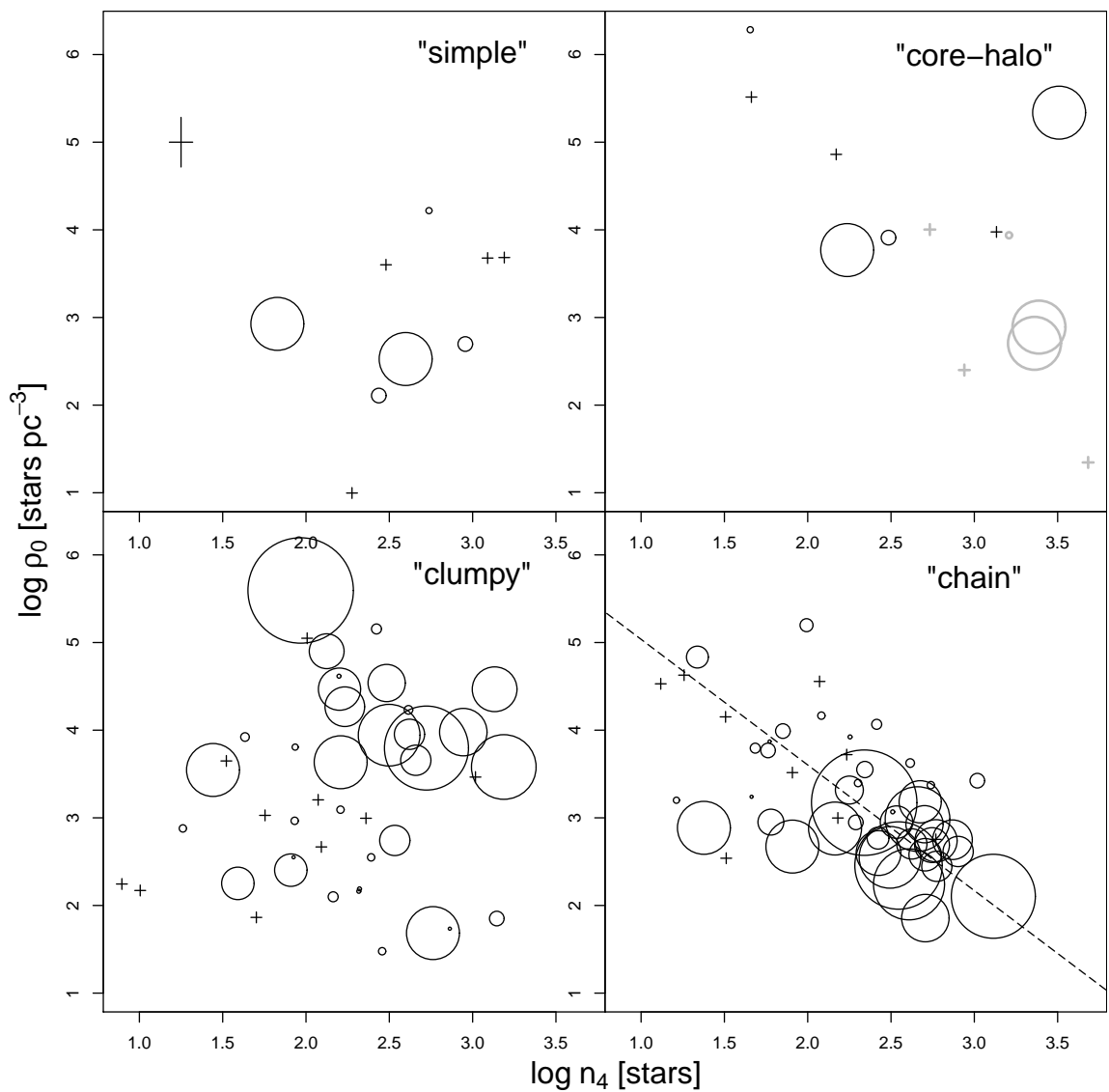}
\caption{Scatter plots of $\rho_0$ vs.\ $n_4$ stratified by the four classes---from left to right and top to bottom: ``simple,'' ``core-halo'' (black indicates ``core''; gray indicates ``halo''), ``clumpy,'' and ``chain.'' Symbol size indicates subcluster age, with smaller circles indicating younger subclusters, larger circles indicating older subclusters, and $+$'s marking subclusters with missing ages. The dashed lines show the reduced major axis regression line for statistically significant relations.
\label{morph2.fig}}
\end{figure}

\clearpage\clearpage

\subsection{Temporal Evolution and Timescales}

We investigate temporal evolution of the principal component of the structural properties ($\log n_4$, $\log r_4$, $\log \Sigma_0$, and $\log \rho_0$) for each morphological class. Using the relations in the definition of these quantities from Equations~\ref{n4.eqn} and \ref{rho0.eqn}, the principal component, a linear combination of any two of these variables, can be expressed as  
\begin{equation}
\log K = C_1\, \log \rho_0 + C_2\, \log r_4 + \mathrm{const.}= (C_1 - C_2/3)\log \rho_0 + (C_2/3)\log n_4+ \mathrm{const.}
\end{equation}
where $C_1$ and $C_2$ are the PCA loadings, which are related to the power-law indices, $\alpha$, of the $\rho_0\sim r_4$ relations from Table~\ref{morph.tab} by
\begin{equation}
C_1 = \sqrt{\frac{1}{\alpha^2+1}}\mathrm{~~~and~~~}C_2 = \sqrt{\frac{\alpha^2}{\alpha^2+1}}.
\end{equation}

The large amounts of scatter in relations between subcluster properties and subcluster ages make it difficult to determine the best functional form when a statistically significant relation is identified. In previous investigations of the evolution of subclusters, both exponential time evolution for gas expulsion \citep[e.g.,][]{Ybarra13} and scale-free time evolution for cluster expansion \citep[e.g.,][]{Gieles12,Pfalzner11} have been explored. We also investigate polynomial regressions. For the regression analysis, we treat age as the independent variable and $K$ as the dependent variable---the functional forms for each regression are shown in Equations~\ref{pl.eqn}--\ref{exp.eqn}, and the regression parameters, $A$, $\alpha$, $N$, and $\tau$
are found using ordinary linear least squares.
\begin{equation}\label{pl.eqn}
\log K = \log A+\alpha\log\mathrm{age}.\textnormal{~~~(power-law)}
\end{equation}
\begin{equation}\label{exp.eqn}
\log K = \log N + 0.434\,\mathrm{age}/\tau.\mathrm{~~~(exponential)}
\end{equation}

Figure~\ref{time_evol.fig} shows the various $K$-variables plotted against age for each sample of subclusters (all subcluster, and stratified by morphological class)---the power-law and exponential regression lines are superimposed. Table~\ref{time.tab} provides the parameters for these fits, and information about the correlations. In all cases a $\log K\sim\mathrm{age}$ has a positive relation, which is statistically significant (using Kendall's $\tau$ test) for samples including of all, ``clumpy,'' and ``chain'' subclusters. The $R^2$ coefficients of determination---a fitting statistic used to determine what fraction of the original scatter is accounted for by the model (i.e., generally $R^2\approx1$ indicate good fits; while $R^2\ll1$ indicate poor fits)---have in a similar range of 0.3--0.5 for both types of models, so we cannot determine whether the power-law or exponential regressions better described the data. However, from the $K$ vs.\ age plots, there may be a hint of upward curvature---particularly for the ``chain'' subcluster sample. 

The exponential curve's $e$-folding timescales vary from 0.5--1~Myr, with the ``clumpy'' subcluster sample having the shortest $\tau=0.56\pm0.14$~Myr timescales, while the ``chain'' subcluster sample having the longest $\tau=1.01\pm0.2$~Myr timescale. This may indicate that dynamical processes proceed faster in clumpy structures, where the subclusters are often overlapping with each other, and more slowly in chain structures, where embedded subcluster are often associated with a discrete molecular core.
Timescales in this range are similar to the 0.8~Myr $e$-folding timescale for gas removal found by \citet{Ybarra13} for molecular clumps in the Rosette Molecular clouds and similar to the 1~Myr age in the numerical models of \citet{Goodwin06} required for a cluster with a 10--30\% star-formation efficiency to double in size.

\begin{deluxetable}{llccl}
\tablecaption{Fits to Structure vs.\ Age Relation \label{time.tab}}
\tabletypesize{\small}\tablewidth{0pt}
\tablehead{
\colhead{Morph.} &  \colhead{Functional} &  \colhead{Parameters} & \colhead{$R^2$}& \colhead{Stat.}\\
\colhead{Class} &  \colhead{Form} & &&  \colhead{Signif.}\\
\colhead{(1)} &  \colhead{(2)} &  \colhead{(3)} &  \colhead{(4)} & \colhead{(5)}
}
\startdata
\vspace{-0.26cm}& power-law & $\log A=-16\pm2$, $\alpha =2.6\pm0.4 $ &  0.32\\
\vspace{-0.26cm}all&&&&$<$0.0001\\
      & exponential & $\log N=-1.1\pm0.2$, $\tau = 0.75\pm0.11\times10^6$ & 0.30\\
 \cline{1-5}
\vspace{-0.26cm} & power-law & $\log A=-13$:, $\alpha = 2.2\pm1.3$ & 0.46 \\
\vspace{-0.26cm}simple&&&&$>$0.05\\
     & exponential & $\log N=-0.6$:, $\tau = 0.87\times10^6$: & 0.37\\
 \cline{1-5}
\vspace{-0.26cm}& power-law & $\log A=-21$:, $\alpha =3.4$: &   0.46\\
\vspace{-0.26cm}core&&&&~~\,0.05\\
      & exponential & $\log N=-1.2\pm0.9$, $\tau = 0.44\times10^6$: & 0.42\\
 \cline{1-5}
\vspace{-0.26cm}& power-law & $\log A=-2.4$:, $\alpha = 3.9$: &   0.37\\
\vspace{-0.26cm}halo&&&&$>$0.05\\
     & exponential & $\log N=-1.7$:, $\tau = 0.45\times10^6$: & 0.29\\
 \cline{1-5}
\vspace{-0.26cm}& power-law & $\log A=-22\pm5$, $\alpha = 3.5\pm0.8$ &   0.36\\
\vspace{-0.26cm}clumpy&&&&$<$0.0001\\
     & exponential & $\log N=-1.5\pm0.4$, $\tau = 0.56\pm0.14$ & 0.32\\
 \cline{1-5}
\vspace{-0.26cm}& power-law & $\log A=-12\pm2$, $\alpha = 1.8\pm0.3$ &   0.36\\
\vspace{-0.26cm}chain&&&&$<$0.0001\\
     & exponential & $\log N=-0.8\pm0.2$, $\tau = 1.01\pm0.20\times10^6$ & 0.38\\
\enddata
\tablecomments{Regression fits to the relation between the principal component of the subcluster structure properties, $K$, and subcluster median age. Column~1: The sample for which $K$ is derived and the regression is performed. Column~2: The equation used to model the $K\sim\mathrm{age}$ relation. Column~3: the values of the regression parameters (using the variable definitions in Equations~\ref{pl.eqn} and \ref{exp.eqn}) and their uncertainties (assuming the functional form) calculated using 1000 iterations of bootstrap resampling. Column~4: The $R^2$ coefficient of determination. Column~5: The null-hypothesis probability, $p(H_0)$, from Kendall's $\tau$ test. Values marked with a colon are highly uncertain.}
\end{deluxetable}

\begin{figure}
\centering
\includegraphics[angle=0.,width=5.0in]{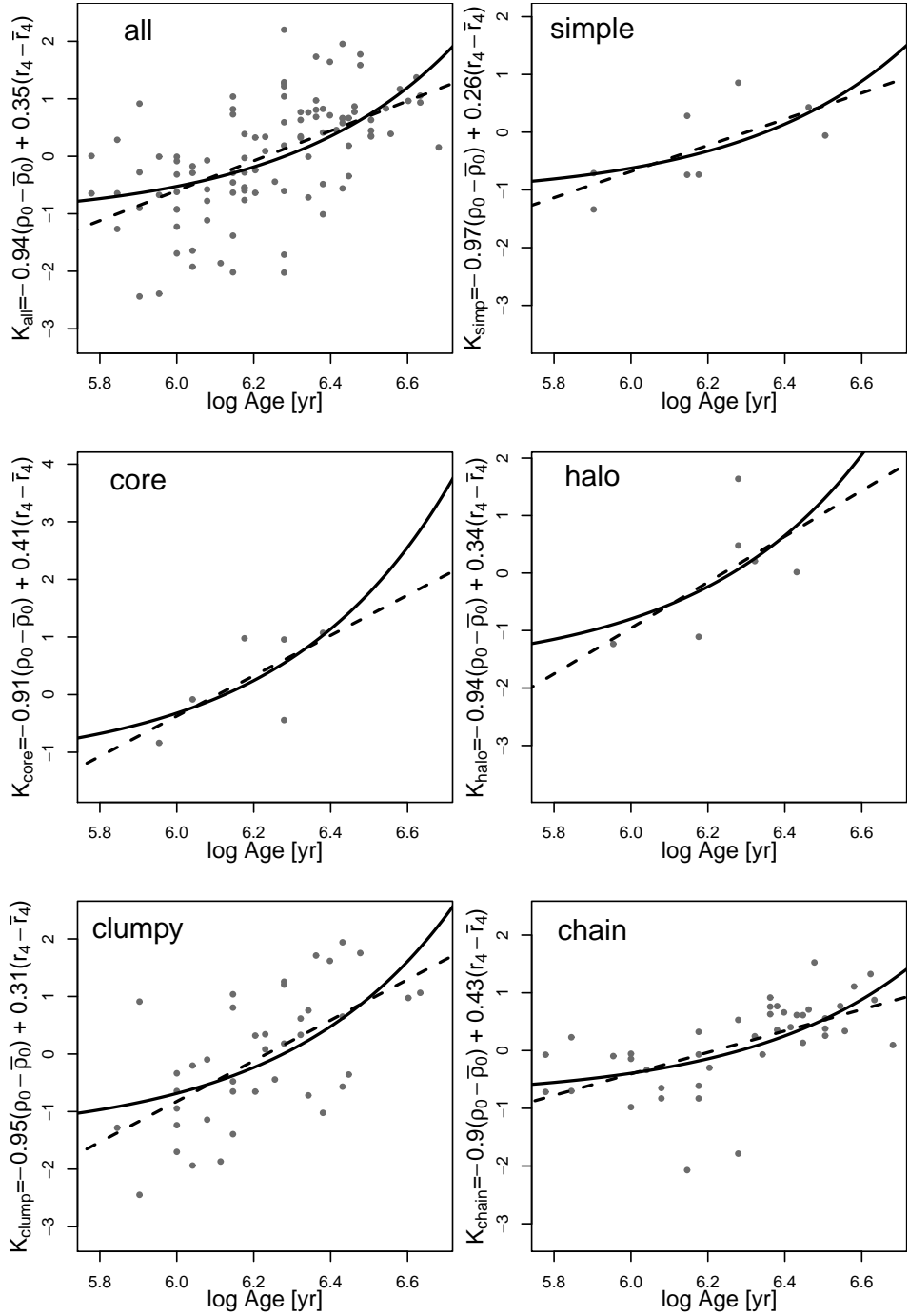}
\caption{Subcluster age vs.\ the principal component, $K$, of the subcluster structure parameters, shown for the sample of all subclusters and subsamples of each morphological class. The y-axes show $K$ computed for each subsample. (The $\rho_0$ and $r_4$ quantities in the y-axis label equations are logarithmic.) Power-law (dashed lines) and exponential (solid lines) regressions are shown.
\label{time_evol.fig}}
\end{figure}

\clearpage\clearpage

\section{Dynamical Subcluster Properties}

A variety of dynamical subcluster properties can be calculated from the physical properties given in Table~\ref{intrinsic.tab}. For these calculations we will ignore the effects of gas mass, considering only the effect of stars down to 0.1~$M_\odot$. The average stellar mass for stars in this mass range from the \citet{Maschberger13} initial mass function is $\bar{m}=0.5~M_\odot$, which we will use to convert from star counts to solar mass units.\footnote{Recall that our definition of $\rho_0$ has units [stars~pc$^{-3}$], not [$M_\odot$~pc$^{-3}$].} 

The free-fall time for a subcluster is
\begin{equation}
t_{ff} = \sqrt{\frac{3\pi}{32\, G\, \bar{m}\, \rho_0}}.
\end{equation}
If a subcluster is in virial equilibrium then the velocity dispersion of the stars will be
\begin{equation}\label{sigmav.eqn}
\sigma_v = \sqrt{ 4\, \pi\, G\, r_c^2\, \bar{m}\,\rho_0 / 9}.
\end{equation}
The crossing time for a cluster requires an assumption for the velocity of stars; two reasonable possibilities include a velocity dispersion of 3~km~s$^{-1}$ as seen by \citet{Furesz06,Furesz08} or the virial velocity dispersion from Equation~\ref{sigmav.eqn}. These two cluster crossing times are
\begin{equation}\label{tcross.eqn}
t_\mathrm{cross,1}\approx r_4 / \mathrm{3~km~s}^{-1}\mathrm{~~or~~}
t_\mathrm{cross,2}\approx r_4 / \sigma_v.
\end{equation}
The subcluster relaxation time will be
\begin{equation}\label{trelax.eqn}
t_\mathrm{relax} \approx t_\mathrm{cross} \frac{n_4}{8 \ln n_4}, 
\end{equation}
and the age of the young stellar cluster in units of $t_\mathrm{relax}$ will be 
\begin{equation}\label{nrelax.eqn}
N_\mathrm{relax} = \mathrm{age} / t_\mathrm{relax}.
\end{equation}
Both Equations~\ref{trelax.eqn} and \ref{nrelax.eqn} require a choice of which formula to use for cluster crossing time. The dynamical quantities obtained from these equations are tabulated in Table~\ref{dynamic.tab} and their univariate distributions are shown in Figure~\ref{dyn_hist.fig}. 

The free-fall times for subclusters tend to be $\sim$1~Myr, slightly younger than the typical age of the subclusters. Thus, if star-formation happens on the one free-fall timescale as postulated by \citet{Elmegreen00}, it should have already ended in most of the subclusters. 

The virial equilibrium velocity dispersions are $<$1~km~s$^{-1}$ which is significantly less than the global velocity dispersions seen in MSFRs like the ONC and NGC~2264. However, gas mass unaccounted for in our derivation of $\sigma_v$ may raise these values somewhat. Given that the survival of the subclusters indicates that they are gravitationally bound, it is unlikely that the true intracluster velocity dispersion is too much larger than the true $\sigma_v$ values when gas mass is accounted for. Cluster crossing times assuming $\sim$3~km~s$^{-1}$ velocities are much less than subcluster ages, so residual structure from star formation should be dynamically erased, on the other hand, this will not be the case assuming velocities similar to $\sigma_v$.

The subcluster relaxation times are mostly much greater than the subcluster ages, which is a common result in studies of young stellar clusters in star-forming regions. Nevertheless, 10\%--40\% of the subclusters have survived for several dynamical timescales. These clusters tend to have both small sizes and small numbers of stars, since both $r_4$ and $n_4$ increase the relaxation time scale in Equations~\ref{tcross.eqn} and \ref{trelax.eqn}. The youngest, embedded clusters often have low values of both $r_4$ and $n_4$, so they often have the shortest dynamical relaxation times. If young stellar clusters are built up by subcluster mergers, the low-$n_4$ clusters may dynamically relax before merging into larger-scale structures, which might inherit the dynamically relaxed state of their composite subclusters \citep[e.g.,][]{Allison09}.

Figure~\ref{trelax_jh.fig} shows the relation between absorption ($J-H$) and relaxation time, indicating that both lightly and heavily absorbed subclusters can have shorter relaxation times, while moderately embedded subclusters have longer relaxation times.

\begin{deluxetable}{lrrrrrrrr}
\tablecaption{Dynamical Subcluster Properties \label{dynamic.tab}}
\tabletypesize{\tiny}\tablewidth{0pt}
\tablehead{
\colhead{Subcluster} & \colhead{$\log t_{ff}$} & \colhead{$\log \sigma_\mathrm{virial}$} & \colhead{$\log t_{\mathrm{cross},1}$} & \colhead{$\log t_{\mathrm{cross},2}$} & \colhead{$\log t_{\mathrm{relax},1}$} & \colhead{$\log t_{\mathrm{relax},2}$} & \colhead{$\log N_\mathrm{relax,1}$} & \colhead{$\log N_\mathrm{relax,2}$} \\
\colhead{} & \colhead{(yr)} & \colhead{(km s$^{-1}$)} & \colhead{(yr)} & \colhead{(yr)} & \colhead{(yr)} & \colhead{(yr)} & \colhead{($t_{\mathrm{relax},1}$)} & \colhead{($t_{\mathrm{relax},2}$)}\\
\colhead{(1)} & \colhead{(2)} & \colhead{(3)} & \colhead{(4)} & \colhead{(5)} & \colhead{(6)}& \colhead{(7)} & \colhead{(8)}& \colhead{(9)}
}
\startdata
orion~A & 4.39 & 0.07 & 4.17 & 4.58 & 4.35 & 4.75 & 1.80 & 1.40\\
orion~B & 5.10 & 0.01 & 4.82 & 5.29 & 5.39 & 5.86 & 0.65 & 0.18\\
orion~C & 5.56 & 0.20 & 5.47 & 5.75 & 6.91 & 7.19 & -0.73 & -1.01\\
orion~D & 5.62 & -0.25 & 5.09 & 5.82 & 5.48 & 6.20 & 0.96 & 0.23\\
flame~A & 5.42 & 0.09 & 5.22 & 5.61 & 6.26 & 6.65 & -0.35 & -0.74\\
w40~A & 5.73 & -0.10 & 5.34 & 5.92 & 6.16 & 6.74 & -0.26 & -0.84\\
rcw36~A & 5.53 & 0.05 & 5.29 & 5.72 & 6.32 & 6.75 & -0.37 & -0.80\\
rcw36~B & 4.77 & -0.05 & 4.43 & 4.96 & 4.61 & 5.14 & \nodata & \nodata\\
ngc2264~A & 5.93 & -0.59 & 5.05 & 6.12 & 4.92 & 5.98 & 1.04 & -0.03\\
ngc2264~B & 5.26 & -0.40 & 4.58 & 5.46 & 4.38 & 5.26 & \nodata & \nodata\\
ngc2264~C & 5.21 & -0.34 & 4.59 & 5.41 & 4.49 & 5.30 & \nodata & \nodata\\
ngc2264~D & 6.08 & -0.59 & 5.21 & 6.28 & 5.19 & 6.25 & 1.32 & 0.25\\
ngc2264~E & 6.09 & -0.33 & 5.48 & 6.28 & 6.04 & 6.85 & 0.46 & -0.34\\
ngc2264~F & 5.45 & -0.33 & 4.83 & 5.65 & 4.90 & 5.71 & \nodata & \nodata\\
ngc2264~G & 5.64 & -0.31 & 5.05 & 5.84 & 5.30 & 6.09 & 0.88 & 0.09\\
ngc2264~H & 6.26 & -0.60 & 5.37 & 6.45 & 5.44 & 6.52 & \nodata & \nodata\\
ngc2264~I & 5.53 & -0.25 & 5.00 & 5.73 & 5.32 & 6.04 & 0.85 & 0.13\\
ngc2264~J & \nodata & \nodata & 5.28 & \nodata & \nodata & \nodata & \nodata & \nodata\\
ngc2264~K & 5.87 & -0.23 & 5.36 & 6.06 & 5.99 & 6.70 & 0.35 & -0.35\\
ngc2264~L & \nodata & \nodata & 4.60 & \nodata & \nodata & \nodata & \nodata & \nodata\\
ngc2264~M & 5.63 & -0.33 & 5.01 & 5.82 & 5.21 & 6.02 & 0.87 & 0.06\\
rosette~A & 7.03 & -0.60 & 6.14 & 7.22 & 6.80 & 7.88 & \nodata & \nodata\\
rosette~B & 6.44 & -0.83 & 5.33 & 6.64 & 5.07 & 6.37 & 1.57 & 0.26\\
rosette~C & \nodata & \nodata & 5.33 & \nodata & \nodata & \nodata & \nodata & \nodata\\
rosette~D & 5.75 & -0.46 & 5.02 & 5.95 & 5.03 & 5.97 & \nodata & \nodata\\
rosette~E & 6.68 & -0.33 & 6.08 & 6.88 & 7.13 & 7.93 & -0.65 & -1.46\\
rosette~F & 6.40 & -0.86 & 5.27 & 6.60 & 4.94 & 6.28 & 1.66 & 0.33\\
rosette~G & \nodata & \nodata & 5.33 & \nodata & \nodata & \nodata & \nodata & \nodata\\
rosette~H & 6.59 & -0.65 & 5.66 & 6.79 & 5.87 & 7.00 & \nodata & \nodata\\
rosette~I & \nodata & \nodata & 5.36 & \nodata & \nodata & \nodata & \nodata & \nodata\\
rosette~J & \nodata & \nodata & 5.22 & \nodata & \nodata & \nodata & \nodata & \nodata\\
rosette~K & \nodata & \nodata & 5.08 & \nodata & \nodata & \nodata & \nodata & \nodata\\
rosette~L & 6.79 & -0.46 & 6.04 & 6.98 & 6.84 & 7.78 & -0.41 & -1.35\\
rosette~M & 6.48 & -0.46 & 5.74 & 6.67 & 6.30 & 7.23 & -0.02 & -0.96\\
rosette~N & \nodata & \nodata & 5.38 & \nodata & \nodata & \nodata & \nodata & \nodata\\
rosette~O & 6.09 & -0.63 & 5.18 & 6.28 & 5.07 & 6.18 & 1.16 & 0.05\\
lagoon~A & 6.25 & -0.46 & 5.51 & 6.45 & 5.88 & 6.82 & 0.46 & -0.48\\
lagoon~B & 5.22 & -0.03 & 4.91 & 5.41 & 5.50 & 6.00 & 0.64 & 0.14\\
lagoon~C & 6.04 & -0.39 & 5.37 & 6.24 & 5.75 & 6.62 & 0.45 & -0.42\\
lagoon~D & 5.70 & -0.41 & 5.01 & 5.90 & 5.08 & 5.97 & 1.17 & 0.28\\
lagoon~E & 6.45 & -0.39 & 5.77 & 6.64 & 6.46 & 7.33 & -0.18 & -1.05\\
lagoon~F & 6.66 & -0.28 & 6.09 & 6.85 & 7.23 & 7.99 & -0.87 & -1.63\\
lagoon~G & 5.56 & -0.33 & 4.95 & 5.76 & 5.11 & 5.92 & 1.23 & 0.43\\
lagoon~H & 6.03 & -0.24 & 5.50 & 6.22 & 6.23 & 6.95 & 0.10 & -0.62\\
lagoon~I & 6.15 & -0.23 & 5.65 & 6.35 & 6.51 & 7.21 & -0.19 & -0.89\\
lagoon~J & 6.19 & -0.39 & 5.52 & 6.39 & 6.03 & 6.89 & 0.40 & -0.46\\
lagoon~K & 6.25 & -0.31 & 5.66 & 6.45 & 6.41 & 7.19 & -0.27 & -1.05\\
ngc2362~A & 6.06 & -0.43 & 5.35 & 6.26 & 5.65 & 6.56 & 0.85 & -0.05\\
ngc2362~B & 6.26 & -0.24 & 5.74 & 6.46 & 6.66 & 7.38 & -0.20 & -0.91\\
dr21~A & \nodata & \nodata & 5.18 & \nodata & \nodata & \nodata & \nodata & \nodata\\
dr21~B & \nodata & \nodata & 4.90 & \nodata & \nodata & \nodata & \nodata & \nodata\\
dr21~C & \nodata & \nodata & 4.89 & \nodata & \nodata & \nodata & \nodata & \nodata\\
dr21~D & 5.57 & -0.12 & 5.16 & 5.76 & 5.80 & 6.40 & 0.05 & -0.55\\
dr21~E & 5.83 & -0.19 & 5.35 & 6.02 & 6.03 & 6.70 & -0.03 & -0.70\\
dr21~F & \nodata & \nodata & 4.74 & \nodata & \nodata & \nodata & \nodata & \nodata\\
dr21~G & \nodata & \nodata & 5.22 & \nodata & \nodata & \nodata & \nodata & \nodata\\
dr21~H & 5.59 & -0.29 & 5.02 & 5.79 & 5.27 & 6.04 & \nodata & \nodata\\
dr21~I & \nodata & \nodata & 5.16 & \nodata & \nodata & \nodata & \nodata & \nodata\\
rcw38~A & 6.69 & -0.18 & 6.39 & 7.05 & 8.24 & 8.90 & -1.96 & -2.62\\
rcw38~B & 4.70 & 0.43 & 5.00 & 5.05 & 6.70 & 6.75 & -0.42 & -0.47\\
rcw38~C & 5.82 & -0.38 & 5.31 & 6.17 & 5.91 & 6.77 & 0.37 & -0.49\\
ngc6334~A & 5.67 & -0.16 & 5.22 & 5.86 & 5.84 & 6.48 & \nodata & \nodata\\
ngc6334~B & 6.18 & -0.20 & 5.69 & 6.37 & 6.63 & 7.31 & -0.27 & -0.95\\
ngc6334~C & 5.25 & -0.08 & 4.89 & 5.44 & 5.38 & 5.93 & \nodata & \nodata\\
ngc6334~D & 5.44 & -0.14 & 5.02 & 5.64 & 5.52 & 6.14 & \nodata & \nodata\\
ngc6334~E & 5.84 & -0.05 & 5.51 & 6.03 & 6.54 & 7.07 & \nodata & \nodata\\
ngc6334~F & 6.03 & -0.30 & 5.44 & 6.22 & 6.02 & 6.80 & \nodata & \nodata\\
ngc6334~G & 5.49 & -0.05 & 5.17 & 5.69 & 5.93 & 6.45 & \nodata & \nodata\\
ngc6334~H & 5.75 & -0.15 & 5.31 & 5.95 & 6.02 & 6.65 & 0.18 & -0.45\\
ngc6334~I & 5.77 & -0.31 & 5.18 & 5.96 & 5.54 & 6.33 & \nodata & \nodata\\
ngc6334~J & 5.81 & 0.05 & 5.58 & 6.01 & 6.85 & 7.28 & -0.68 & -1.11\\
ngc6334~K & 5.91 & -0.43 & 5.19 & 6.10 & 5.37 & 6.28 & \nodata & \nodata\\
ngc6334~L & 5.99 & -0.18 & 5.53 & 6.19 & 6.38 & 7.03 & -0.53 & -1.19\\
ngc6334~M & \nodata & \nodata & 5.38 & \nodata & \nodata & \nodata & \nodata & \nodata\\
ngc6334~N & \nodata & \nodata & 5.58 & \nodata & \nodata & \nodata & \nodata & \nodata\\
ngc6357~A & 5.69 & 0.12 & 5.52 & 5.88 & 6.86 & 7.22 & -0.71 & -1.07\\
ngc6357~B & 6.18 & -0.09 & 5.80 & 6.37 & 7.02 & 7.59 & -0.88 & -1.45\\
ngc6357~C & 5.79 & 0.06 & 5.57 & 5.99 & 6.84 & 7.26 & -0.76 & -1.18\\
ngc6357~D & 4.95 & 0.14 & 4.81 & 5.14 & 5.58 & 5.92 & 0.46 & 0.12\\
ngc6357~E & \nodata & \nodata & 5.78 & \nodata & \nodata & \nodata & \nodata & \nodata\\
ngc6357~F & 5.69 & 0.15 & 5.55 & 5.88 & 6.97 & 7.30 & -0.80 & -1.12\\
eagle~A & 5.64 & -0.15 & 5.21 & 5.84 & 5.83 & 6.46 & 0.55 & -0.08\\
eagle~B & 6.17 & 0.04 & 5.94 & 6.37 & 7.50 & 7.94 & -1.18 & -1.62\\
eagle~C & 5.92 & -0.30 & 5.34 & 6.12 & 5.83 & 6.61 & 0.40 & -0.38\\
eagle~D & 6.60 & -0.17 & 6.15 & 6.80 & 7.53 & 8.18 & -1.13 & -1.78\\
eagle~E & \nodata & \nodata & 5.01 & \nodata & \nodata & \nodata & \nodata & \nodata\\
eagle~F & 6.43 & -0.39 & 5.76 & 6.63 & 6.45 & 7.32 & \nodata & \nodata\\
eagle~G & 6.01 & -0.44 & 5.29 & 6.21 & 5.54 & 6.45 & \nodata & \nodata\\
eagle~H & \nodata & \nodata & 5.11 & \nodata & \nodata & \nodata & \nodata & \nodata\\
eagle~I & 6.33 & -0.49 & 5.55 & 6.52 & 5.92 & 6.88 & -0.01 & -0.98\\
eagle~J & 6.40 & -0.62 & 5.50 & 6.60 & 5.62 & 6.72 & \nodata & \nodata\\
eagle~K & \nodata & \nodata & 4.93 & \nodata & \nodata & \nodata & \nodata & \nodata\\
eagle~L & \nodata & \nodata & 4.61 & \nodata & \nodata & \nodata & \nodata & \nodata\\
m17~A & 5.29 & -0.05 & 4.96 & 5.49 & 5.55 & 6.08 & \nodata & \nodata\\
m17~B & 4.73 & 0.06 & 4.51 & 4.92 & 4.92 & 5.33 & \nodata & \nodata\\
m17~C & 5.55 & 0.00 & 5.27 & 5.74 & 6.21 & 6.68 & -0.06 & -0.54\\
m17~D & 5.74 & 0.13 & 5.59 & 5.93 & 7.00 & 7.35 & -0.96 & -1.31\\
m17~E & 5.39 & -0.07 & 5.04 & 5.59 & 5.66 & 6.21 & 0.72 & 0.17\\
m17~F & 5.08 & -0.00 & 4.79 & 5.27 & 5.32 & 5.80 & \nodata & \nodata\\
m17~G & 5.00 & -0.02 & 4.70 & 5.20 & 5.14 & 5.64 & \nodata & \nodata\\
m17~H & 5.41 & 0.05 & 5.18 & 5.61 & 6.11 & 6.54 & -0.11 & -0.54\\
m17~I & 5.63 & 0.01 & 5.36 & 5.82 & 6.38 & 6.85 & -0.24 & -0.70\\
m17~J & \nodata & \nodata & 4.16 & \nodata & \nodata & \nodata & \nodata & \nodata\\
m17~K & 5.54 & 0.12 & 5.37 & 5.73 & 6.58 & 6.94 & -0.58 & -0.94\\
m17~L & 5.29 & 0.26 & 5.27 & 5.49 & 6.64 & 6.86 & -0.56 & -0.78\\
m17~M & 5.70 & -0.03 & 5.38 & 5.89 & 6.35 & 6.86 & \nodata & \nodata\\
m17~N & 5.56 & -0.04 & 5.24 & 5.75 & 6.07 & 6.59 & 0.13 & -0.38\\
m17~O & 5.26 & 0.06 & 5.03 & 5.45 & 5.86 & 6.28 & -0.01 & -0.43\\
carina~A & 5.71 & -0.19 & 5.24 & 5.90 & 5.84 & 6.50 & 0.61 & -0.06\\
carina~B & 6.08 & 0.08 & 5.88 & 6.28 & 7.48 & 7.87 & -1.04 & -1.44\\
carina~C & 5.54 & 0.18 & 5.44 & 5.73 & 6.81 & 7.11 & -0.63 & -0.93\\
carina~D & 6.24 & -0.20 & 5.76 & 6.43 & 6.77 & 7.44 & -0.39 & -1.06\\
carina~E & 6.05 & -0.19 & 5.58 & 6.25 & 6.44 & 7.11 & -0.06 & -0.73\\
carina~F & 6.41 & -0.29 & 5.84 & 6.60 & 6.76 & 7.53 & -0.18 & -0.95\\
carina~G & \nodata & \nodata & 6.77 & \nodata & \nodata & \nodata & \nodata & \nodata\\
carina~H & 5.94 & -0.11 & 5.55 & 6.13 & 6.53 & 7.11 & -0.08 & -0.67\\
carina~I & 5.94 & -0.22 & 5.44 & 6.13 & 6.14 & 6.84 & 0.54 & -0.16\\
carina~J & 6.22 & -0.12 & 5.81 & 6.41 & 6.99 & 7.59 & -0.63 & -1.23\\
carina~K & 6.03 & -0.14 & 5.61 & 6.23 & 6.58 & 7.20 & -0.02 & -0.64\\
carina~L & 6.15 & -0.11 & 5.76 & 6.35 & 6.90 & 7.49 & -0.47 & -1.06\\
carina~M & 6.18 & -0.16 & 5.74 & 6.38 & 6.78 & 7.42 & -0.38 & -1.02\\
carina~N & 6.15 & -0.15 & 5.72 & 6.34 & 6.78 & 7.41 & \nodata & \nodata\\
carina~O & 5.71 & -0.05 & 5.38 & 5.91 & 6.31 & 6.84 & -0.27 & -0.80\\
carina~P & 6.47 & -0.14 & 6.05 & 6.67 & 7.41 & 8.02 & -0.79 & -1.40\\
carina~Q & 6.30 & -0.27 & 5.75 & 6.49 & 6.62 & 7.37 & 0.01 & -0.73\\
carina~R & 6.60 & -0.32 & 6.00 & 6.79 & 7.01 & 7.80 & -0.53 & -1.32\\
carina~S & 6.23 & -0.29 & 5.66 & 6.42 & 6.43 & 7.20 & 0.03 & -0.74\\
carina~T & 6.30 & -0.20 & 5.83 & 6.50 & 6.89 & 7.56 & -0.53 & -1.20\\
trifid~A & \nodata & \nodata & 6.05 & \nodata & \nodata & \nodata & \nodata & \nodata\\
trifid~B & 5.57 & -0.05 & 5.24 & 5.77 & 6.07 & 6.59 & 0.21 & -0.31\\
trifid~C & 6.33 & -0.15 & 5.90 & 6.52 & 7.10 & 7.73 & -0.82 & -1.45\\
trifid~D & 6.47 & -0.36 & 5.83 & 6.67 & 6.61 & 7.45 & -0.33 & -1.17\\
ngc1893~A & 6.25 & -0.27 & 5.70 & 6.45 & 6.53 & 7.28 & 0.01 & -0.73\\
ngc1893~B & 6.06 & -0.14 & 5.64 & 6.26 & 6.65 & 7.26 & -0.23 & -0.85\\
ngc1893~C & 6.19 & -0.45 & 5.46 & 6.39 & 5.82 & 6.75 & 0.68 & -0.24\\
ngc1893~D & 5.11 & -0.28 & 4.55 & 5.30 & 4.50 & 5.25 & 1.78 & 1.03\\
ngc1893~E & \nodata & \nodata & 5.08 & \nodata & \nodata & \nodata & \nodata & \nodata\\
ngc1893~F & 6.05 & -0.44 & 5.33 & 6.25 & 5.59 & 6.51 & 0.73 & -0.19\\
ngc1893~G & 6.05 & -0.27 & 5.50 & 6.25 & 6.16 & 6.91 & 0.01 & -0.74\\
ngc1893~H & 6.14 & -0.26 & 5.60 & 6.34 & 6.37 & 7.11 & -0.09 & -0.83\\
ngc1893~I & 6.16 & -0.15 & 5.73 & 6.35 & 6.79 & 7.41 & -0.34 & -0.97\\
ngc1893~J & 4.93 & 0.00 & 4.65 & 5.12 & 5.08 & 5.55 & 1.07 & 0.60\\
\enddata
\tablecomments{Dynamical properties of individual ellipsoidal subclusters. Column~1: Subcluster designation. Column~2: Free-fall time for the subcluster (stellar mass only). Column~3: Mean velocity dispersion for stars within 4 core radii assuming that the subcluster is virialized. Column~4: Subcluster crossing time assuming a velocity dispersion of 3.0~km~s$^{-1}$. Column~5: Subcluster crossing time assuming a virial velocity dispersion. Column~6: Dynamical relaxation time assuming a velocity dispersion of 3.0~km~s$^{-1}$. Column~7: Dynamical relaxation time assuming a virial velocity dispersion. Column~8--9: Subcluster dynamical age in relaxation times.
}
\end{deluxetable}

\begin{figure}
\centering
\includegraphics[angle=0.,width=4.4in]{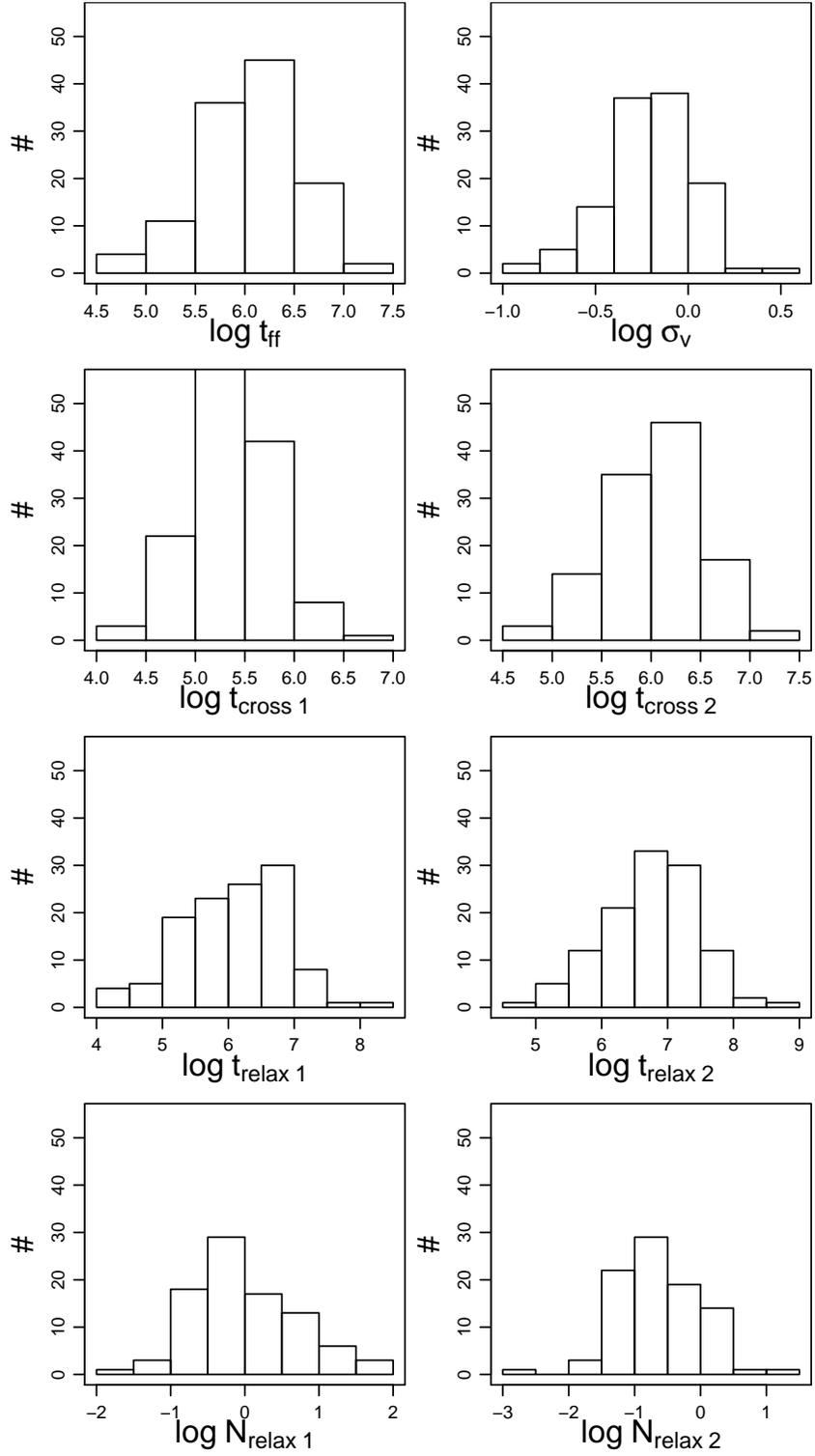}
\caption{Univariate histograms of dynamical properties of subclusters. Top to bottom and left to right: $t_{ff}$, $\sigma_v$, $t_\mathrm{cross,1}$, $t_\mathrm{cross,2}$, $t_\mathrm{relax,1}$, $t_\mathrm{relax,2}$, $N_\mathrm{relax,1}$, and $N_\mathrm{relax,2}$.
\label{dyn_hist.fig}}
\end{figure}

\begin{figure}
\centering
\includegraphics[angle=0.,width=3.6in]{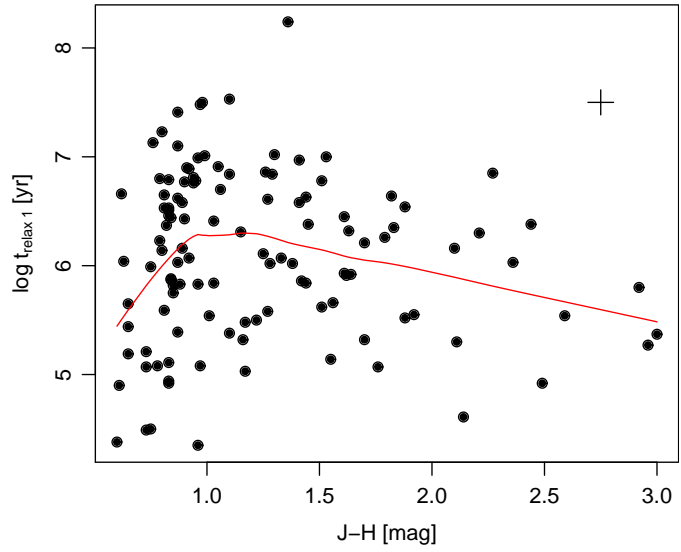}
\caption{Scatter plot of $t_\mathrm{relax,1}$ vs $J-H$, with a LOWESS line plotted. (The relation using $t_\mathrm{relax,2}$ is similar.)
\label{trelax_jh.fig}}
\end{figure}

\clearpage\clearpage



\section{Conclusions}

Our primary empirical results and their implications for star-formation theory are listed below.

\begin{enumerate}

\item Subclusters with larger ages have larger radii, which we validate using Kendall's $\tau$ test ($p<10^{-4}$). We interpret this to be a sign of subcluster expansion, which was postulated by \citet{Tutukov78} and others as a consequence of the loss of molecular-cloud material due to star-formation feedback. This helps confirm the inference of cluster expansion by \citet{Pfalzner11} from the density--radius relation in the catalog of young stellar clusters from \citet{Lada03}. 

\item The inferred expansion rate, which for most subclusters is a radial increase less than 1~km~s$^{-1}$, is lower than the radial velocities observed in many spectroscopic studies of young stellar clusters \citep[e.g.,][]{Furesz06,Furesz08,Tobin15} or generated by simulations \citep[e.g.,][]{Bate03}. However, other clusters may have slower stellar velocity dispersions \citep[e.g.,][]{Jeffries14}. If the MYStIX regions have velocity dispersions $\ge$3~km~s$^{-1}$ the subclusters would need to be gravitationally bound for their structure to be preserved. However, for clusters that have undergone expansion, $\sim$1~km~s$^{-1}$ expansion rates are consistent with the decrease in velocity dispersion seen in young stellar cluster N-body simulations \citep{Moeckel10}.
This result has implications for the model of cluster survival by \citet{Kruijssen12} because the fraction of stars that form in naturally bound subclusters is an intermediate step in determining what fraction of stars will remain in bound clusters after the end of star formation. Their model can be applied before the end of star formation in a region, so, by combining the \mystix\ results with additional studies of the molecular clouds in these regions, the cluster-survival model can be tested on Galactic star-forming regions.

\item We observe the strong correlation from \citet{Pfalzner11} between cluster radius and cluster density. (There is a slight difference in the quantities being measured because they use total cluster radius while we use subcluster core radius.) We observe a similar trend, $\rho_0 \propto r_4^{-2.3}$, a track slightly less steep than would be expected for expansion from a uniform initial state with a conserved number of stars. The flatter slope indicates that one of those conditions must not hold. \citet{Pfalzner11} explains this phenomenon as an effect of star formation progressing from the inner part of the clusters, first, to the outer part of the cluster, later, (the inside--out scenario). However, the age gradients found by \citet{Getman14b} show the opposite trends, ruling out the inside--out scenario. We note that hierarchical cluster mergers also offer an explanation for this effect because, as clusters have time to expand, they have time to merge with subclusters; and mergers with subclusters can cause expansion due to conservation of energy. 

\item Larger subcluster central volume density, $\rho_0$, is correlated with smaller numbers of stars, $n_4$ ($p<0.05$; marginally significant)---which is consistent with our observation of a flat $\Sigma_0 \sim n_4$ relation. This effect, combined with the distribution of moderately elliptical subclusters from Paper~I \citep[cf.][]{Maschberger10} and the less steep $\rho_0 \sim r_4$ relation, suggests that young stellar clusters grow from hierarchical subcluster mergers \citep[e.g.,][]{McMillan07}. If subcluster mergers are an important aspect of early young stellar cluster dynamical evolution, it would lead to more rapid dynamical relaxation \citep{Allison09,Parker12}, which could be used to explain unexpected results, like the mass segregation seen in W~40 \citep[eg.,][]{Kuhn10} or well-fit isothermal sphere profiles from Paper~I.

\item The morphological classifications of structure of star-forming regions from Paper~I (``simple,'' ``core-halo,'' ``clumpy,'' ``chain'') are not correlated with age. This may be considered evidence against an evolutionary progression. However, an alternate explanation for the lack of a trend could be differences in rates of dynamical evolution in different environments. Gas expulsion is expected to slow mergers of subclusters \citep[e.g.,][]{Kruijssen12}, so when ``chain'' clusters---which are highly substructured suggesting an early stage of evolutionary progression \citep{Parker14}---lose their gas, as in the case of NGC~1893, they may maintain their initial structure inherited from the molecular cloud as they age.

\item When comparing the subcluster age and the number of stars in a subcluster for all \mystix\ regions simultaneously, there is no evidence of a trend. However, this may be influenced by differing cloud environments, as we suggest for the morphological classes. When we control for absorption, selecting only the most absorbed subclusters for $n_4\sim \mathrm{age}$ analysis, the number of stars does increase with radius ($p<0.05$; marginally significant). Thus, the age--number of stars relation is also consistent with a picture of subcluster mergers.

\item Two-body relaxation times calculated for subclusters tend to be much longer than the age of the subclusters. However, a fraction of the subclusters with low-$n_4$ and low-$r_4$, have had time to dynamically relax. This is an important aspect of models of accelerated dynamical relaxation through cluster mergers \citep{Allison09,Parker12}.


\end{enumerate}

Finally, it is important to note that the analysis of correlations in cluster properties performed in this paper is sensitive to the way in which subclusters are defined, which is true for the Paper~I subcluster catalog as well as the \citet{Lada03} catalog and the \citet{Gutermuth09} catalog. For the type of analysis performed here, the \mystix\ subcluster catalog has several advantages over these other catalogs. First, the \mystix\ MPCM catalog contains relatively large samples of low-mass young stars (both disk-bearing and disk-free) without containing too many non-cluster member contaminants \citep{Feigelson13,Broos13}. Second, the subcluster analysis avoids combining distinct clumps of stars into artificial clusters, and instead focuses on the properties of subclusters. Third, parametric modeling disentangles subcluster properties like density and radius, instead of using a surface-density threshold to define subcluster boundaries, which would lead to subclusters with higher surface density having artificially larger radii (Paper~I).

\acknowledgements Acknowledgments:  


The \mystix\ project was supported at Penn State by NASA grant NNX09AC74G, NSF grant AST-0908038, and the \Chandra\ ACIS Team contract SV4-74018 (G.~Garmire \& L.~Townsley, Principal Investigators), issued by the \Chandra\ X-ray Center, which is operated by the Smithsonian Astrophysical Observatory for and on behalf of NASA under contract NAS8-03060.  M.A.K. also received support from NSF SI2-SSE grant AST-1047586 (G. J. Babu, PI), a CONICYT Gemini grant (J. Borissova, PI) from the Programa de Astronom\'{i}a del DRI Folio 32130012 (Fondecyt Regular No.\ 1120601), and the Chilean Ministry of Economy, Development, and Tourism's Millennium Science Initiative through grant IC12009,  awarded to Millennium Institute of Astrophysics, and Fondecyt project number 3150319. J.B. is supported by FONDECYT No.1120601. We thank Susanne Pfalzner, Leisa Townsley, Patrick Broos, G. Jogesh Babu, Kevin Luhman, and Richard Wade for providing advice on this paper. We also thank the anonymous referee for comments on the manuscript. This research made use of data products from the \Chandra\ Data Archive. This work is based on observations made with the {\it Spitzer Space Telescope}, obtained from the NASA/IPAC Infrared Science Archive, both of which are operated by the Jet Propulsion Laboratory, California Institute of Technology under a contract with the National Aeronautics and Space Administration. This research has also made use of SAOImage DS9 software developed by Smithsonian Astrophysical Observatory and NASA's Astrophysics Data System Bibliographic Services.

\end{document}